\documentclass[11pt,a4paper,twoside,groupcitations]{article}
\pdfoutput=1
\usepackage[T1]{fontenc}
\usepackage[ansinew]{inputenc}
\usepackage[italian,english]{babel}
\usepackage{amsfonts}
\usepackage{amsmath}
\usepackage{array}
\usepackage{amsthm}
\usepackage{amssymb}
\pdfoutput=1
\usepackage{graphicx}
\usepackage{subfigure}
\usepackage{braket}
\usepackage{eucal}
\usepackage{verbatim}
\usepackage[table]{xcolor}
\usepackage{caption}
\usepackage{multicol}
\usepackage{tikz}
\usetikzlibrary{arrows,shapes}
\usetikzlibrary{matrix,arrows}
\newcommand{\be}{\begin{eqnarray}}
\newcommand{\ee}{\end{eqnarray}}

\newcommand{\expec}[1]{\mbox{$\langle\, #1\,\rangle$}}

\renewcommand{\d}{\mbox{${\rm d}$}} 
\newcommand{\lp}{\ell_{\rm p}}
\newcommand{\mpl}{m_{\rm p}}
\newcommand{\gn}{G_{\rm N}}
\newcommand{\gd}{G_{D}}
\newcommand{\rh}{r_{\rm H}}

\newcommand{\RS}{R_{\rm S}}
\newcommand{\RSD}{R_{D}}

\newcommand{\E}{\mathrm{E}}
\title{\bf Horizon of quantum black holes in various dimensions}
\author{Roberto~Casadio$^{ab}$\thanks{E-mail: casadio@bo.infn.it},
$\ $
Rogerio~T.~Cavalcanti$^c$\thanks{E-mail: rogerio.cavalcanti@ufabc.edu.br},
$\ $
Andrea~Giugno$^{ab}$\thanks{E-mail: andrea.giugno2@unibo.it},
$\ $
and
Jonas~Mureika$^{d}$\thanks{E-mail: jmureika@lmu.edu}
\\
\\
{\em $^a$Dipartimento di Fisica e Astronomia, Universit\`a di Bologna}
\\
{\em via Irnerio~46, I-40126 Bologna, Italy}
\\
\\
{\em $^b$I.N.F.N., Sezione di Bologna,}
\\
{\em via B.~Pichat~6/2, I-40127 Bologna, Italy}
\\
\\
{\em $^c$Centro de Ci\^{e}ncias Naturais e Humanas, Universidade Federal do ABC}
\\
{\em 09210-580, Santo Andr\'e - SP, Brazil}
\\
\\
{\em $^d$Department of Physics, Loyola Marymount University}
\\
{\em 1 LMU Drive, Los Angeles, CA, USA 90045}
}
\begin{document}
\maketitle
\begin{abstract}
%
We adapt the horizon wave-function formalism to describe massive static spherically
symmetric sources in a general $(1+D)$-dimensional space-time, for $D>3$ and
including the $D=1$ case.
We find that the probability $P_{\rm BH}$ that such objects are (quantum) black holes 
behaves similarly to the probability in the $(3+1)$ framework for $D> 3$.
In fact, for $D\ge 3$, the probability increases towards unity as the mass grows above
the relevant $D$-dimensional Planck scale $m_D$.
At fixed mass, however, $P_{\rm BH}$ decreases with increasing $D$,
so that a particle with mass $m\simeq m_D$ has just about $10\%$ probability
to be a black hole in $D=5$, and smaller for larger $D$.
This result has a potentially strong impact on estimates of black hole production in
colliders.
In contrast, for $D=1$, we find the probability is comparably larger for smaller masses, but
$P_{\rm BH} < 0.5$, suggesting that such lower dimensional black holes are purely quantum
and not classical objects.
This result is consistent with recent observations that sub-Planckian 
black holes are governed by an effective two-dimensional gravitation theory.
Lastly, we derive Generalised Uncertainty Principle relations for the black holes
under consideration, and find a minimum length corresponding to a
characteristic energy scale of the order of the fundamental gravitational mass $m_D$
in $D>3$.
For $D=1$ we instead find the uncertainty due to the horizon fluctuations has the same
form as the usual Heisenberg contribution, and therefore no fundamental scale exists.
\end{abstract}

\section{Introduction}
\setcounter{equation}{0}
\label{intro}
Unusual causal structures like trapping surfaces and horizons
can only occur in strongly gravitating systems, such as astrophysical
objects that collapse and possibly form black holes. 
One might argue that for a large black hole, 
gravity should appear “locally weak” at the horizon, 
since tidal forces look small to a freely falling observer 
(their magnitude being roughly controlled by the surface gravity, 
which is inversely proportional to the horizon radius). 
Like any other classical signal, light
is confined inside the horizon no matter how weak such forces 
may appear to a local observer.  This can be taken as
the definition of a ``globally strong'' interaction. 
\par
As the black hole's mass approaches the Planck scale, 
tidal forces become strong both in the local and global sense,
thus granting such an energy scale a remarkable role in the search
for a quantum theory of gravity. 
It is indeed not surprising that modifications 
to the standard commutators of quantum mechanics 
and Generalised Uncertainty Principles (GUPs) have been proposed,
essentially in order to account for the possible existence 
of small black holes around the Planck scale, 
and the ensuing minimum measurable length~\cite{Hossenfelder:2012jw}.
Unfortunately, that regime is presently well beyond our experimental capabilities, 
at least if one takes the Planck scale at face value~\footnote{We use units where
$c=1$ and $\hbar=\lp\,\mpl=\ell_D\,m_D$.}, $\mpl\simeq 10^{16}\,$TeV
(corresponding to a length scale $\lp=\hbar/\mpl=\mpl\,\gn\simeq10^{-35}\,$m). 
Nonetheless, there is the possibility that the low energy theory still retains some
signature features that could be accessed in the near future 
(see, for example, Refs.~\cite{Hogan:2015kva}).
\subsection{Gravitational radius and horizon wave-function}
Before we start calculating phenomenological predictions,
it is of the foremost importance
that we clarify the possible conceptual issues arising from the use of arguments and
observables that we know work at our every-day scales. 
One of such key concepts is the gravitational radius of a self-gravitating source,
which can be used to assess the existence of trapping surfaces, 
at least in spherically symmetric systems.
As it is very well known, the location of a trapping surface 
is determined by the equation
\be
g^{ij}\nabla_i r \nabla_j r
\,=\,
0
\ ,
\ee
where $\nabla_i r$ is perpendicular to surfaces of constant area
$\mathcal{A} = 4\pi r^2$. 
If we set $x^1 = t$ and $x^2 = r$, 
and denote the matter density as $\rho=\rho(r,t)$, 
the Einstein field equations tell us that
\be
g^{rr}
\,=\,
1-\frac{2\lp(m/\mpl)}{r}
\ ,
\ee
where the Misner-Sharp mass is given by
\be
m(r,t)
\,=\,
4\pi \int_0^r \rho(\bar{r},t)\bar{r}^2 \, \d \bar{r}
\ ,
\ee
as if the space inside the sphere were flat. 
A trapping surface then exists if there are values of $r$ 
and $t$ such that the gravitational radius
$
\RS
=
2\,\lp \, {m}/{\mpl}
\ ,
$
satisfies
\be
\RS(r,t)
\,\geq\,
r
\ .
\ee
When the above relation holds in the vacuum outside the region 
where the source is located, $\RS$ becomes the usual Schwarzschild radius, 
and the above argument gives a mathematical
foundation to Thorne’s hoop conjecture~\cite{Thorne:1972ji}, 
which (roughly) states that a black hole forms 
when the impact parameter $b$ of two colliding small objects is shorter than the Schwarzschild
radius of the system, that is for $b\lesssim 2\,\lp\, E/\mpl$
where $E$ is the total energy in the centre-of mass frame.
\par
If we consider a spin-less point-particle of mass $m$, 
the Heisenberg principle of quantum mechanics introduces an uncertainty 
in the particle’s spatial localisation of the order of the Compton
scale $\lambda_m \simeq \lp\,\mpl/m$~\footnote{Strictly speaking, this bound
holds in the non-relativistic limit $E\lesssim 2\,m$~\cite{dodonov}, but we shall employ it
in this work since we always consider particles and black holes in their rest frame.}.
Since quantum physics is a more refined description of reality, we
could argue that $\RS$ only makes sense if~\footnote{One could also
derive this condition from the famous Buchdahl's inequality~\cite{buchdahl}, which is however
a result of classical general relativity, whose validity in the quantum domain we cannot
take for granted.}
\be
\RS
\,\gtrsim\,
\lambda_m
\quad\Longrightarrow\quad
m
\,\gtrsim\,
\mpl
\ ,
\ee
which brings us to face the conceptual challenge of describing quantum mechanical
systems whose classical horizon would be smaller than the size of the uncertainty
in their position.
In Refs.~\cite{Casadio:2013tma}, a proposal was put forward
in order to describe the ``fuzzy'' Schwarzschild
(or gravitational) radius of a localised but likewise fuzzy quantum source.
One starts from the spectral decomposition of the spherically symmetric
wave-function
\be
\Ket{\psi_{\rm S}}
=
\sum_{E} C(E) \Ket{\psi_{E}}
\ ,
\ee
with the usual constraint
\be
\hat{H} \Ket{\psi_{E}}
=
E\Ket{\psi_{E}}
\ ,
\ee
and associates to each energy level $\Ket{\psi_{\rm E}}$ a probability amplitude
$\psi_{\rm H}(\RS)\simeq C(E)$, where $\RS=2\,\lp\,E/\mpl$.
From this Horizon Wave-Function (HWF), a GUP and minimum measurable 
length were derived~\cite{Casadio:2013aua}, as well as corrections to the classical
hoop conjecture~\cite{Casadio:2013uga}, and a modified time evolution
proposed~\cite{Casadio:2014twa}. 
The same approach was generalised to electrically charged sources~\cite{RN}, 
and used to show that Bose-Einstein condensate models of
black holes~\cite{DvaliGomez,Casadio:2015bna,Kuhnel:2014oja,mueckPT,Hofmann:2014jya}
actually possess a horizon with a proper semiclassical limit~\cite{Casadio:2014vja}.
\par
It is important to emphasise that the HWF approach differs 
from most previous attempts in which the gravitational degrees of freedom 
of the horizon, or of the black hole metric, are
quantised independently of the nature and state of the source 
(for some bibliography, see, e.g., Ref.~\cite{Davidson:2014tda}). 
In our case, the gravitational radius is instead quantised along with the
matter source that produces it, somewhat more in line 
with the highly non-linear general relativistic description of the gravitational interaction.
However, having given a practical tool for describing 
the gravitational radius of a generic quantum system 
is just the starting point.
In fact, when the probability that the source is localised within 
its gravitational radius is significant, the system should show (some of) 
the properties ascribed to a black hole in general relativity. 
These properties, the fact in particular that no signal can escape from
the interior, only become relevant once we consider how the overall system evolves. 
\subsection{Higher and lower dimensional models}
Extra-dimensions have been proposed as a possible explanation for
some of the incongruences affecting particle physics, such as the hierarchy
problem between fundamental interactions.
In $(1+D)$-dimensional space-times, with $D\geq4$, gravity shows its true quantum
nature at a scale $m_D$ (possibly much) lower than the Planck mass $\mpl$.
Such scenarios have been extensively studied after the well known
ADD~\cite{ArkaniHamed:1998rs} and
Randall-Sundrum~\cite{Randall:1999ee} models
were proposed (see Ref.~\cite{Maartens:2010ar} for a comprehensive review).
However our purpose is not to study any model in particular, but to see how
the probability of a microscopic black hole formation could be affected by
assuming the existence of extra dimensions.
We shall therefore just consider black holes in the ADD scenario with
a horizon radius significantly shorter than the size of the extra dimensions. 
It is then important to recall that in these models the Newton constant
is replaced by the gravitational constant
\be
G_D
=
\frac{\ell_D^{D-2}}{m_D}
\ ,
\label{GD}
\ee
where $\ell_D=\hbar/m_D\gg\lp$ is the new gravitational length scale.
\par
On the other hand, gravitational theories become much simpler
in space-times with fewer than 3 spatial dimensions,
where corresponding quantum theories are exactly solvable~\cite{Mureika:2012na}.
Such theories have been revisited in recent years, motivated by
model-independent evidence that the number of space-time dimensions 
decreases as the Planck length is approached.
Such formalisms -- known generically as ``spontaneous dimension reduction''
mechanisms --  have been studied
in various contexts, mostly focusing on the energy-dependence
of the space-time spectral dimension, including causal dynamical
triangulations~\cite{cdt1} and non-commutative geometry inspired
mechanisms~\cite{lmpn,Nicolini:2012fy,Carr:2015nqa,Mureika:2012fq}.  
An alternative approach suggests the effective dimensionality of
space-time increases as the ambient energy scale
drops~\cite{vd1,jrmds1,nads,Stojkovic:2014lha}.
\par
Given these arguments, we will generealize the results of
Ref.~\cite{Casadio:2014twa} in an arbitrary number of spatial
dimensions.
In Section~\ref{GaussModes} we will introduce the concept horizon
wave-function and we will apply it to a system described
by a gaussian wave-packet. 
Subsequently, we will compute the probability that the system
is a black hole in Section~\ref{Probability} and obtain a 
Generalised Uncertainty Principle in Section~\ref{secGUP}.
Finally we will give some conclusions and possible outlook about
the obtained results in Section~\ref{Conclusions}.
\section{Static horizon-wave function in higher dimensions}
\setcounter{equation}{0}
\label{GaussModes}
We recall that, given any spherically symmetric function $f=f(r)$ in $D$ spatial dimensions,
the corresponding function in momentum space is given by
\be
\tilde{f}(p)
\!\!&=&\!\!
 \frac{p^{\frac{2-D}{2}}}{\ell_D\,m_D}\int_0^\infty \d r \, r^{D/2}\,  J_{\frac{D-2}{2}}\left(\frac{r \,p}{\ell_D\,m_D}\right)f(r)
\ ,
\label{Fourier}
\ee
where the normalised radial modes are given by the Bessel functions
\be
J_{\frac{D-2}{2}}\left(\frac{r \,p}{\ell_D\,m_D}\right)
=
\frac{\Omega_{D-2}}{(2\pi)^{D/2}} \, \left(\frac{r \,p}{\ell_D\,m_D}\right)^{\frac{D-2}{2}}
\int_0^\pi \d\theta\, e^{-i\,p\,r\,\cos\theta /\ell_D\,m_D}\,(\sin \theta)^{D-2}
\ ,
\ee
and, accordingly, the inverse transform is given by
\be
{f}(r)
=
\frac{r^{\frac{2-D}{2}}}{\ell_D\,m_D}
\int_0^\infty \d p
\,
p^{D/2}\,
J_{\frac{D-2}{2}}\left(\frac{r\, p}{\ell_D\,m_D}\right)
\tilde{f}(p)
\ .
\ee
\par
We can apply the above definitions to a localised massive particle described by the
Gaussian wave-function
\be
\psi_{\rm S}(r)
=
\frac{e^{-\frac{r^2}{2\,\ell^2}}}{(\ell\, \sqrt{\pi})^{D/2}}
\ ,
\label{Gauss}
\ee
and the corresponding function in momentum space is thus given by
\be
\tilde{\psi}_{\rm S}(p)
=
\frac{e^{-\frac{p^2}{2\,\Delta^2}}}{(\Delta\, \sqrt{\pi})^{D/2}}
\ ,
\label{momGauss}
\ee
where $\Delta=m_D \,\ell_D/\ell$ is the spread of the wave-packet in momenta space.

If we assume
\be
\ell
\geq
\lambda_m
\equiv
\frac{\hbar}{m}
=
\frac{m_D\,\ell_D}{m}
\ ,
\ee
where $\lambda_m$ is the Compton wavelength, the smallest resolvable scale
associated to the particle according to quantum mechanics~\cite{dodonov}, It immediately
follows that $\Delta\leq m$.
Note that we have
\be
\frac{\ell}{\lambda_m}
=
\frac{m}{\Delta}
\ ,
\label{lCompt}
\ee
which will allow us to express $\Delta$ in terms of $\ell$.
\subsection{$(1+D)$-dimensional Schwarzschild metric}
The Schwarzschild metric, as a solution of the vacuum Einstein equations,
generalises in $(1+D)$-dimensional space-time as
\be
ds^2
=
-\left(1-\frac{\RSD}{r^{D-2}}\right) \, \d t^2
+\left(1-\frac{\RSD}{r^{D-2}}\right)^{-1} \, \d r^2
+r^{D-1} \, \d\Omega_{D-1}
\ ,
\ee
where the classical horizon radius is given by
\be
\RSD
=
\left(\frac{2 \, \gd \, M}{|D-2|}\right)^{\frac{1}{D-2}} 
=
\begin{cases} 
\frac{1}{2 \,G_1\, M} & \quad \text{if }\ D=1
\\
&
\\
\left(\frac{2 \,\gd\, M}{D-2}\right)^{\frac{1}{D-2}}
& \quad \text{if }\ D > 2 
\\
\end{cases}
\ .
\label{SchwD}
\ee
Of course, if $D=3$ we have the standard result $R_{3}=\RS$.
We note that the $D=1$ limit of the
horizon radius given above matches that obtained from an exact solution to 
Einstein's equations in $(1+1)$-dimensions \cite{mann}.
We are also purposefully avoiding $(1+2)$-dimensional models,
such as the BTZ black holes, because they have meaning only in anti-de Sitter
space-time and we are not dealing with a cosmological constant.
\par
As in $D=3$, we assume the relativistic mass-shell relation in flat
space~\cite{Casadio:2013tma},
\be
p^2
=
E^2-m^2
\ ,
\ee
and define the HWF expressing the energy $E$ of the particle
in terms of the related horizon radius~\eqref{SchwD}, $\rh=R_D(E)$.
From Eq.~\eqref{momGauss}, we then get
\be
\psi_{\rm H} (\rh)
\!\!&=&\!\!
\mathcal{N}_{\rm H}\,
\Theta(\rh-R_D)
\, \exp\left\{ -
\frac{1}{2}\left(\frac{D-2}{2\,\gd\, \Delta}\right)^2
\, \left[ \rh^{2\,(D-2)} - R_D^{2\,(D-2)}
\right] \right\}
\ .
\label{nnormHWF}
\ee
The normalisation $\mathcal{N}_{\rm H}$ is fixed according to
\be
\mathcal{N}_{\rm H}^{-2} e^{-m^2/\Delta^2}
\!\!&=&\!\!
\frac{\Omega_{D-1}}{\mathcal{N}_{\rm H}^2} \, e^{-m^2/\Delta^2} \,
\int_0^\infty |\psi_{\rm H}(\rh)|^2 \, \rh^{D-1} \, \d \rh
\notag
\\
\!\!&=&\!\!
\frac{\pi^{D/2}}{D-2} \, \left(\frac{2\gd \Delta}{D-2}\right)^{\frac{D}{D-2}}
\, \frac{\Gamma\left(\frac{D}{2D-4},\frac{m^2}{\Delta^2}\right)}{\Gamma\left(\frac{D}{2}\right)}
\ ,
\ee
where
\be
\Gamma(s,x)
=
\int_x^\infty t^{s-1} \, e^{-t} \, \d t
\ee
is the upper incomplete Euler Gamma function,
and, using Eq.~\eqref{SchwD}, we obtain
\be
\psi_{\rm H}(\rh)
\!\!&=&\!\!
\left\{
\frac{D-2}{\ell_D^{\,D} \, \pi^{D/2}}
\left[\frac{(D-2)\,m_D}{2\,\Delta}\right]^{\frac{D}{D-2}}
\, \frac{\Gamma\left(\frac{D}{2}\right)}
{\Gamma\left(\frac{D}{2D-4},\frac{m^2}{\Delta^2}\right)}
\right\}^{1/2}
\nonumber
\\
&&
\times
\Theta(\rh-R_D)\, \exp\left\{ -
\frac{(D-2)^2}{8} \, \frac{m_D^2}{\Delta^2}
\, \left(\frac{\rh}{\ell_D}\right)^{2(D-2)} \right\}
\ ,
\label{HWF}
\ee
for $D>2$.
\par
Finally, we remark that these results also hold in $D=1$,
although with a significant change in the step function.
In fact, according to~\eqref{SchwD}, the condition $E\gtrsim M$
now yields $\rh\lesssim R_1$, and the HWF reads
\be
\psi_{\rm H}(\rh)
\,=\,
\sqrt{\frac{2/\ell}
{\Gamma\left(-\frac{1}{2},\frac{m^2}{\Delta^2}\right)}}
\, \Theta(R_1-|\rh|)\, \exp\left\{ -
\frac{\ell^2}{8\rh^2} \right\}
\ ,
\label{HWF1}
\ee
where we also used $G_1\, \Delta=1/\ell$.
\section{Black hole probability}
\setcounter{equation}{0}
\label{Probability}
According to standard definitions, the probability density that the Gaussian particle
lies inside its own gravitational radius is the following product
\be
\mathcal{P}_<(r<\rh)
\,=\,
P_{\rm S}(r<\rh) \, \mathcal{P}_{\rm H}(\rh)
\ ,
\label{PInside}
\ee
where the probability that the particle is inside a $D-$ball of radius $\rh$ is
\be
P_{\rm S}(r<\rh)
=
\Omega_{D-1}\int_0^{\rh} |\psi_{\rm S}(r)|^2 \, r^{D-1} \, \d r
\ ,
\ee
and the probability density that the radius of the horizon equals $\rh$ is
\be
\mathcal{P}_{\rm H}(\rh)
=
\Omega_{D-1} \, r^{D-1} \, |\psi_{\rm H}(\rh)|^2
\ee
Integrating~\eqref{PInside} over all the possible values of the horizon radius $\rh$, 
\be
P_{\rm BH}
\,=\,
\int_0^\infty
\mathcal{P}_<(r<\rh) \, \d \rh
\label{PBH}
\ee
gives the probability that the particle is a black hole.
\subsection{Higher-dimensional space-times}
\begin{figure}[t!]
\centering
\raisebox{4.5cm}{ $P_{\rm BH}$}
\includegraphics[width=6.5125cm,height=5cm]{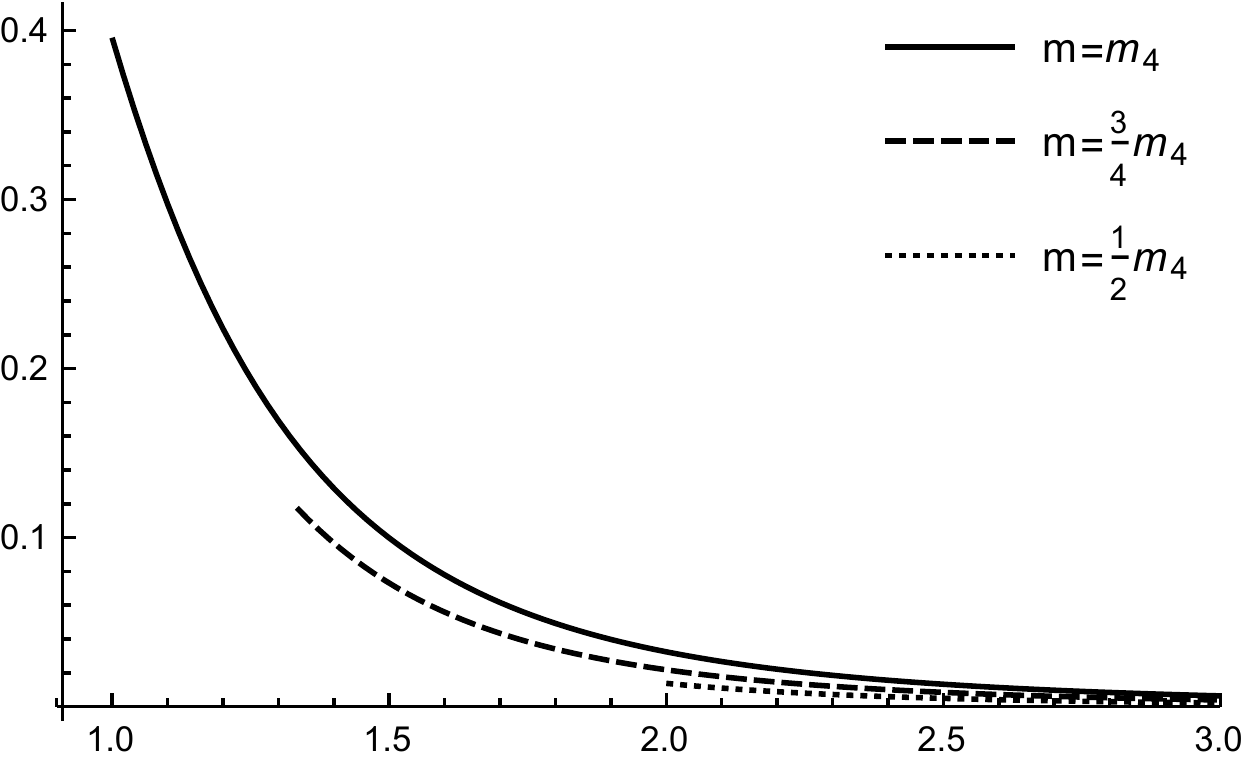}
$\quad$
\includegraphics[width=6.5125cm,height=5cm]{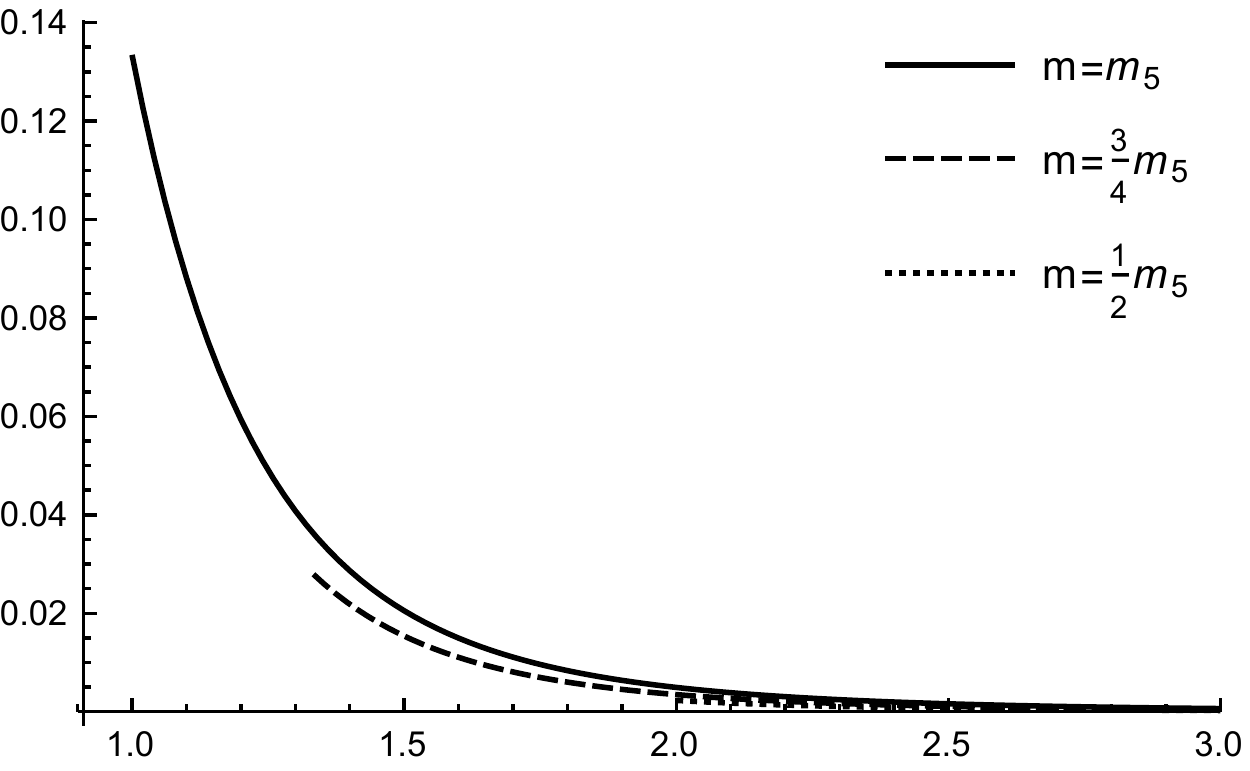}
{$\frac{\ell}{\ell_D}$}
\\
$D=4$
\hspace{6cm}
$D=5$
\\
\raisebox{4.5cm}{ $P_{\rm BH}$} 
\includegraphics[width=6.5125cm,height=5cm]{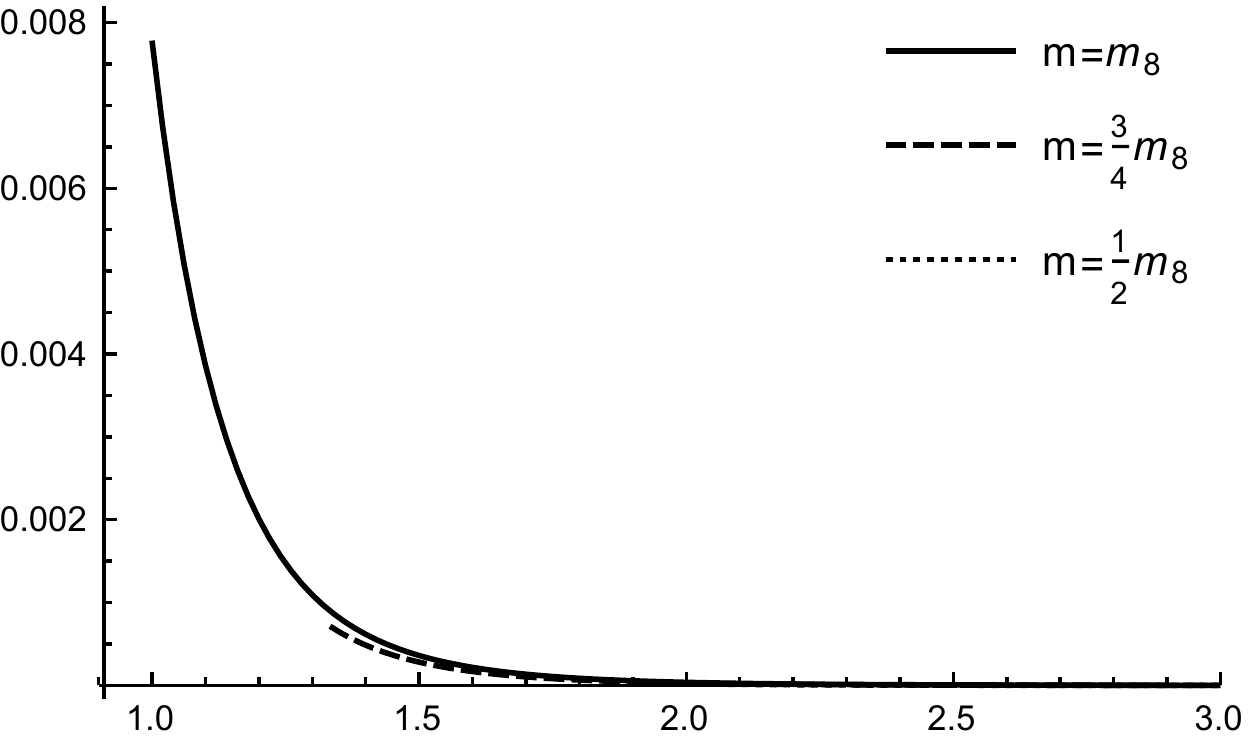}
$\quad$
\includegraphics[width=6.5125cm,height=5cm]{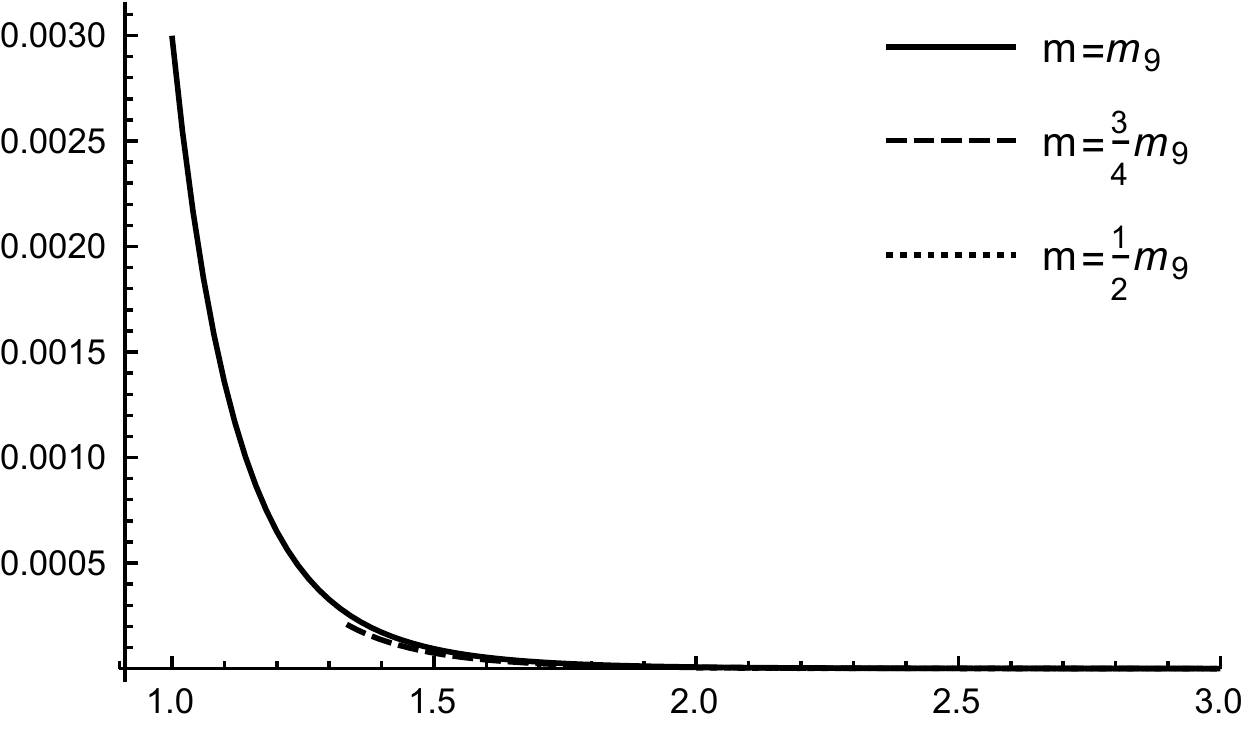}
{$\frac{\ell}{\ell_D}$}
\\
$D=8$
\hspace{6cm}
$D=9$
\caption{Probability $P_{\rm BH}(\ell,m)$ of a particle, described by a 
Gaussian with $\ell\ge\lambda_m$, to be a black hole for $m=m_D$
(solid line), $m=3\,m_D/4$ (dashed line) and $m=m_D/2$ (dotted line).
From left to right, the spatial dimensions are $D=4$ and 5 on the first line
and $D=8$ and 9 on the second line
(note the different scales on the vertical axes). 
\label{prob1}}
\end{figure}

Now we can employ the results of Section~\ref{Gauss}.
First, we have
\be
P_{\rm S}(r<\rh)
=
\frac{\gamma\left(\frac{D}{2},\frac{\rh^2}{\ell^2}\right)}{\Gamma\left(\frac{D}{2}\right)}
\ ,
\ee
where
$\gamma$ is the lower incomplete Gamma function
\be
\gamma(s,x)
=
\Gamma(s)-\Gamma(s,x)
\ .
\ee
Properties of the $\gamma$ ensure that $P_{\rm S}=1$ if $\rh\to\infty$, 
while $P_{\rm S}=0$ if $\rh=0$.
Then,
\be
\mathcal{P}_{\rm H}(\rh)
\!\!&=&\!\!
\frac{2}{\ell_D^{\,D}}\, \left(\frac{(D-2) \, m_D}{2 \, \Delta}\right)^{\frac{D}{D-2}}
\frac{D-2}{\Gamma\left(\frac{D}{2\,D-4},\frac{m^2}{\Delta^2}\right)}
\nonumber
\\
&&
\times
\Theta(\rh-R_D) \, \exp\left\{ -
\left[\frac{(D-2)\, m_D}{2 \, \Delta}\right]^2
\, \left(\frac{\rh}{\ell_D}\right)^{2(D-2)} \right\}
\, \rh^{D-1}
\ .
\ee
and Eq.~\eqref{PBH} finally becomes
\be
P_{\rm BH}
\!\!&=&\!\!
\frac{2}{\ell_D^{\,D}}\left(\frac{(D-2)\, m_D}{2\,\Delta}\right)^{\frac{D}{D-2}}
\frac{D-2}{\Gamma\left(\frac{D}{2D-4},\frac{m^2}{\Delta^2}\right)
\Gamma\left(\frac{D}{2}\right)}
\notag
\\
&&
\times
\int_{R_D}^\infty
\gamma\left(\frac{D}{2},\frac{\rh^2}{\ell^2}\right)
 \exp\left\{ -
\left[\frac{(D-2)\, m_D}{2 \, \Delta}\right]^2
\, \left(\frac{\rh}{\ell_D}\right)^{2(D-2)} \right\}
\, \rh^{D-1} \, \d \rh
\ ,
\label{PBHexplicit}
\ee
which yields the probability for a particle to be a black hole depending
on the Gaussian width $\ell$, mass $m$ and spatial dimension $D$.
Since the above integral cannot be performed analytically, we show
the numerical dependence on $\ell\gtrsim \lambda_m$~\footnote{We
recall that one expect $\ell$ is bounded from below by the Compton
length of the source.}
of the above probability for different masses and spatial dimensions
in Fig.~\ref{prob1}.
\par
One immediately notices that the probability $P_{\rm BH}$
at given $m$ decreases significantly for increasing $D$, and for large
values of $D$ even a particle of mass $m\simeq m_D$ is most likely not
a black hole.
This result should have a strong impact on predictions of black hole
production in particle collisions.
For example, one could approximate the effective production cross-section
as $\sigma(E)\sim P_{\rm BH}(E)\,\sigma_{\rm BH}(E)$, where
$\sigma_{\rm BH}\sim 4\,\pi\,E^2$ is the usual black disk expression for a
collision with centre-of-mass energy  $E$.
Since $P_{\rm BH}$ can be very small, $\sigma(E)\ll \sigma_{\rm BH}(E)$
for $D>4$ and one in general expects much less black holes can be produced
than standard estimates~\cite{landsberg}.
\par
\begin{figure}[h!]
\centering
\raisebox{2cm}{\tiny $P_{\rm BH}$}
\includegraphics[width=3.3cm,height=2.625cm]{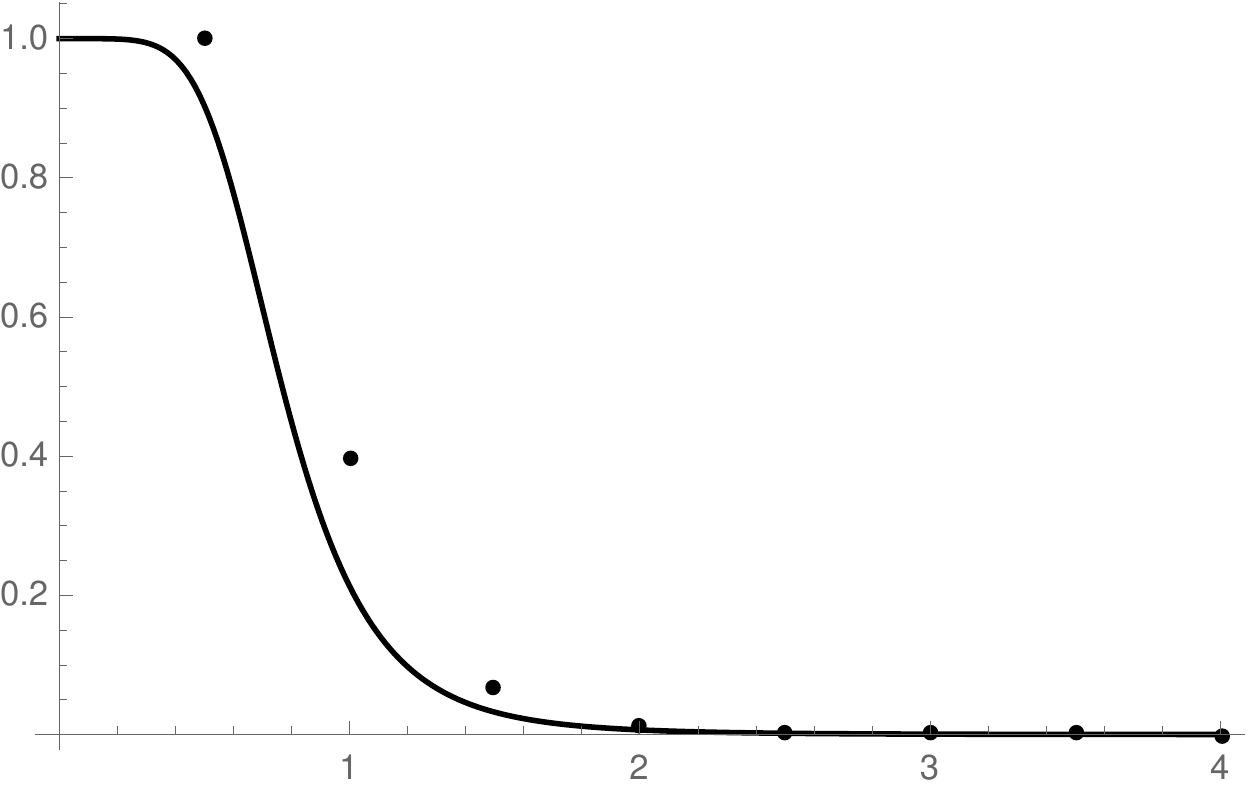}
$\quad$
\includegraphics[width=3.3cm,height=2.625cm]{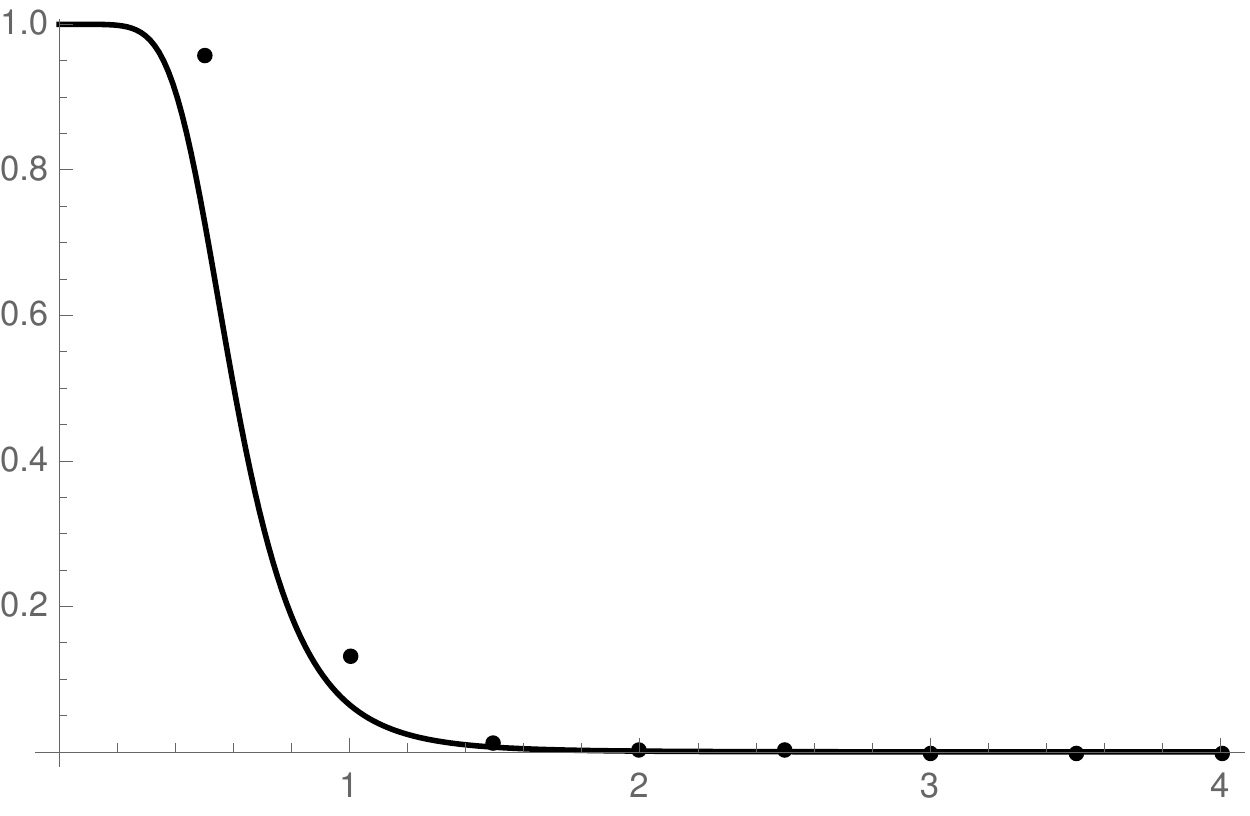}
$\quad$
\includegraphics[width=3.3cm,height=2.625cm]{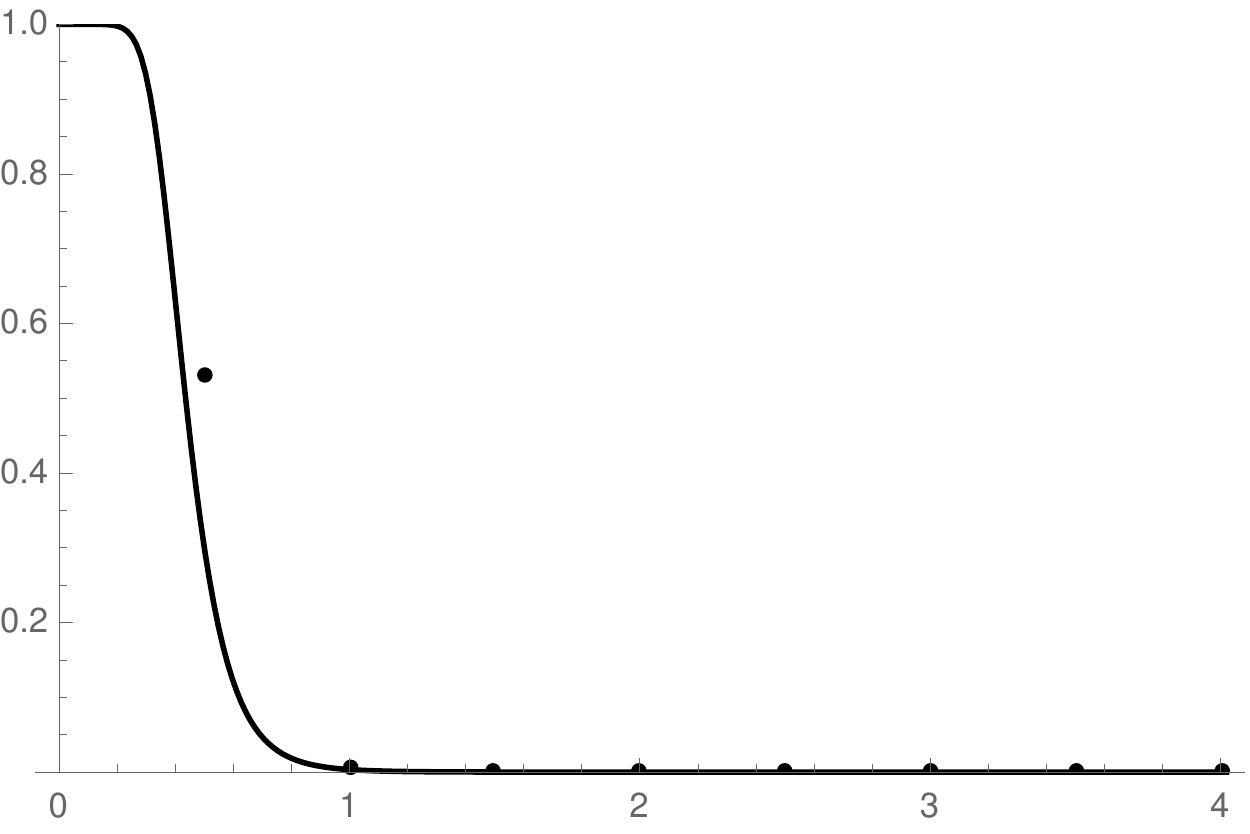}
$\quad$
\includegraphics[width=3.3cm,height=2.625cm]{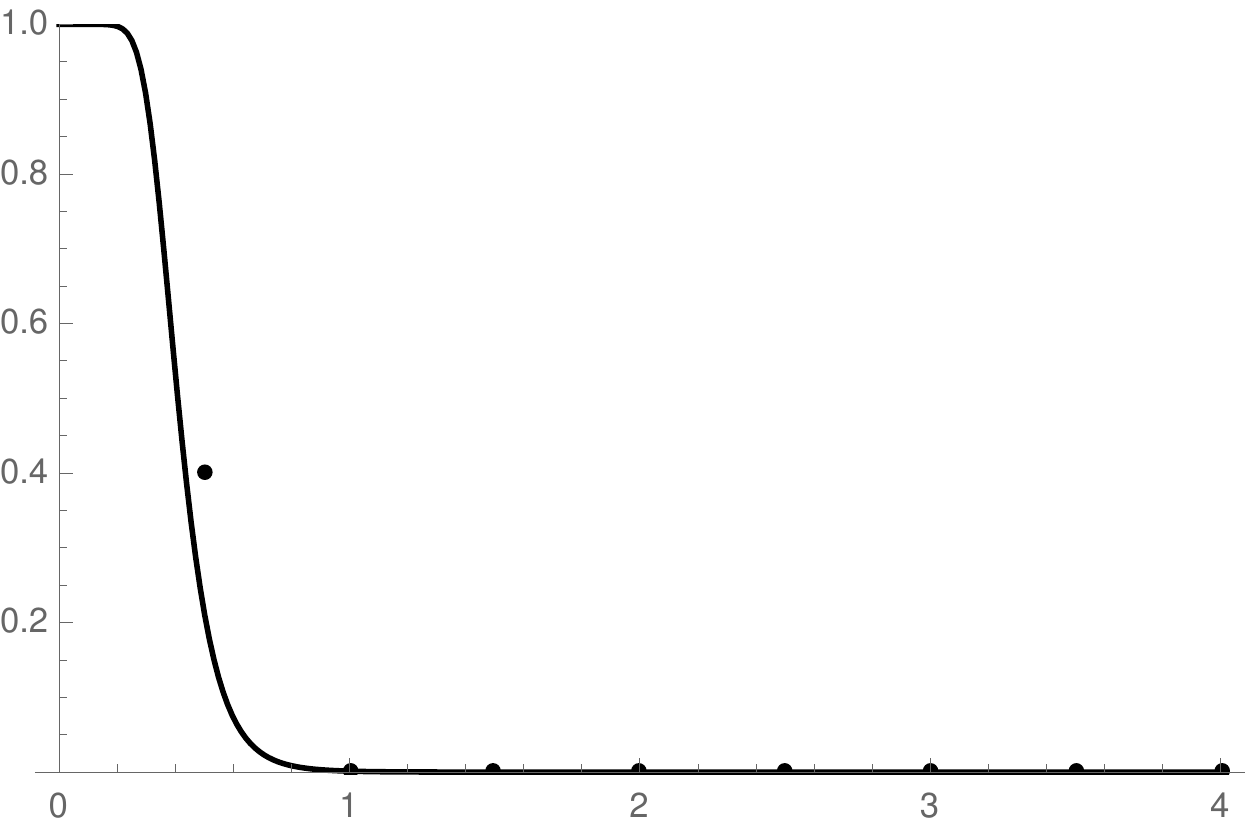}
{\tiny $\frac{\ell}{\ell_D}$}
\\
$D=4$
\hspace{3cm}
$D=5$
\hspace{3cm}
$D=8$
\hspace{3cm}
$D=9$
\caption{Probability $P_{\rm BH}(\ell)$ of a particle, described by a 
Gaussian with $\ell=\lambda_m$, to be a black hole (dots)
compared to its analytical approximation.
\label{prob2}}
\end{figure}

For the particular case $\ell=\lambda_m$, which according to the Eq.~\eqref{lCompt}
implies $m=\Delta$, the probability depends only on $\ell$ and $D$.
Further, the expression~\eqref{PBHexplicit} can be approximated analytically
by taking the limit $R_D\to 0$.
Fig.~\ref{prob2} compares this approximation with the numerical results,
showing that it describes fairly well the correct behaviour~\footnote{Note that
we include values of $m\gg m_D$ (corresponding to $\ell\ll\ell_D$) in order
to obtain large probabilities.}.
\subsection{$(1+1)-$dimensional space-time}
\begin{figure}[t]
\centering
\raisebox{4.5cm}{ $P_{\rm BH}$}
\includegraphics[width=6.5125cm,height=5cm]{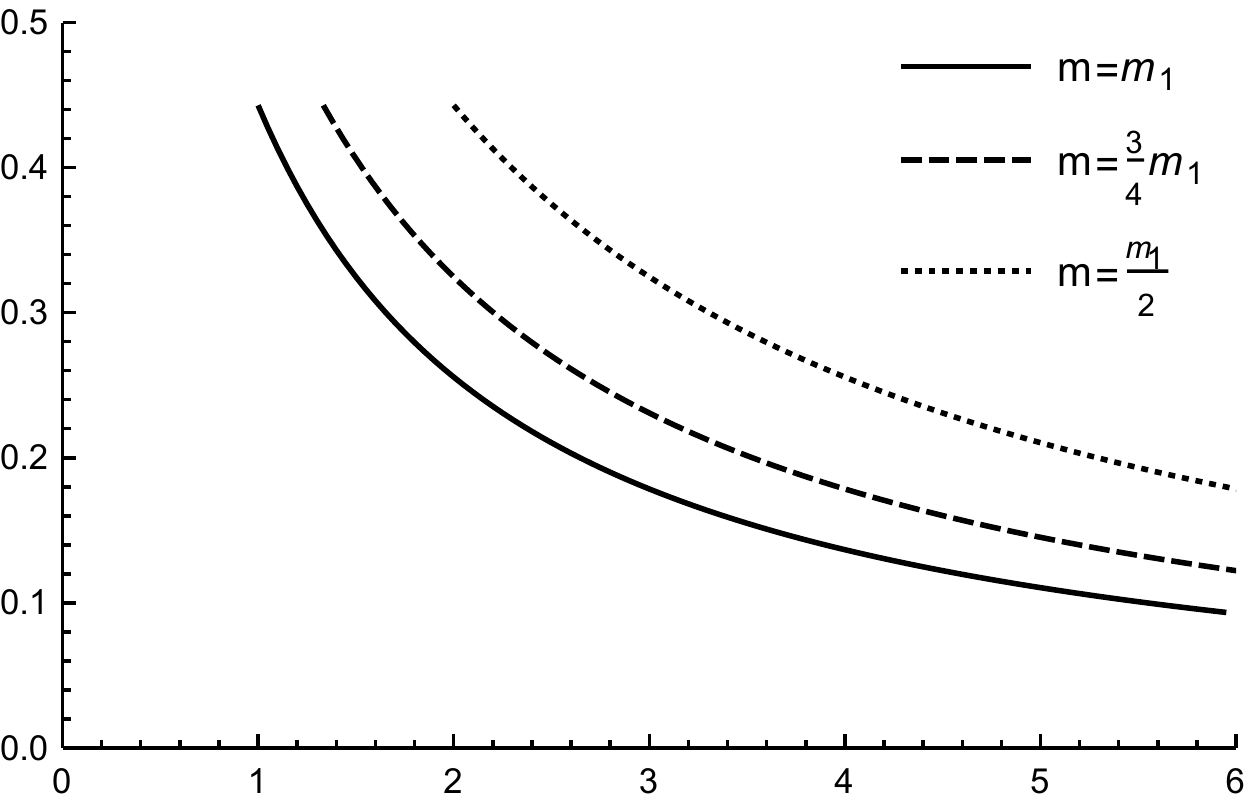}
$\quad${ $\ell/\ell_1$}
\caption{Probability $P_{\rm BH}(\ell,m)$ for a particle to be
a black hole in $D=1$, for $m=m_1$ (solid line), $m=3\,m_1/4$ (dashed line)
and $m=m_1/2$ (dotted line).
\label{prob3}}
\end{figure}

In $D=1$, from Eqs.~\eqref{Gauss} and~\eqref{HWF1}, we have
\be
P_{\rm S}(r<\rh)
=
\mathrm{erf}\left(\frac{\rh}{\ell}\right)
\ ,
\ee
and
\be
\mathcal{P}_{\rm H}(\rh)
=
\frac{2/\ell}
{\Gamma\left(-\frac{1}{2},\frac{m^2}{\Delta^2}\right)}
\, \Theta(R_1-|\rh|)\, \exp\left\{ -
\frac{\ell^2}{4\,\rh^{2}}\right\}
\ .
\ee
Consequently, the black hole probability is
\be
P_{\rm BH}
=
\frac{4/\ell}
{\Gamma\left(-\frac{1}{2},\frac{m^2}{\Delta^2}\right)}
\int_0^{R_1}
\mathrm{erf}\left(\frac{\rh}{\ell}\right)
\exp\left\{ -\frac{\ell^2}{4\,\rh^{2} }\right\} \, \d \rh
\ .
\ee
We note that this formula can just be obtained from Eq.~\eqref{PBHexplicit}
by setting $D=1$ and taking the complementary integration domain. 
Fig.~\ref{prob3} shows this probability as a function of $\ell\ge \lambda_m$
for various masses.
\par
An immediate feature that differentiates the $(1+1)$-dimensional case from the 
$(1+D)$-dimensional cases is the rate of increase of the probability for source
mass less than the Planck scale $m_1$.
In higher dimensions, the probability of forming a black hole with $m \geq m_D$
is quite high for $\ell\simeq \ell_D$, and drops significantly, at the
same length scale, for $m < m_D$.
For $D=1$, the drop is much slower and sources with $m < m_1$ have 
a comparably larger probability to be black holes. 
However, the main difference is that the maximum $P_{\rm BH}\simeq 0.45$,
precisely obtained for $\ell= \lambda_m$, and does not depend on $m$.
This is in agreement with the fact that, in $D=1$, the gravitational
constant $G_1=1/\hbar$ and
\be
\expec{\hat r_{\rm H}}
\simeq 
R_1(m)
\simeq
\lambda_m
\ ,
\ee
for any values of $m$.

This means the larger the mass $m$, the smaller the Compton length and
the horizon radius (which would instead be larger in $D\ge 3$).
Correspondingly, for a given width $\ell$, the probability the object is a black
hole increases for decreasing mass (according to the fact that $\expec{\hat r_{\rm H}}$
becomes larger), and the source can never be a truly classical black hole
(with $\expec{\hat r_{\rm H}}\gg \lambda_m$) in $D=1$. 

\par
This result is consistent with the notion that black holes in $(1+1)$-dimensional
space-time are strictly quantum objects, as discussed in Ref.~\cite{Mureika:2012fq}.  
Furthermore, it can be understood to support several recent results suggesting the
gravitational physics (and corresponding black holes) in the 
sub-Planckian regime is two dimensional \cite{Nicolini:2012fy,Carr:2015nqa}.
Said another way, black holes in the (sub-)Planckian
regime will naturally form as effective $(1+1)$-dimensional objects, and in this
sense the duality in mass-dependence between the $(1+3)$- and $(1+1)$-dimensional
Schwarzschild metric is due to dimensional reduction.
\section{GUP from HWF}
\label{secGUP}
\begin{figure}[h!]
\centering
\raisebox{4.5cm}{$\frac{\expec{\hat{r}_{\rm H}}}{R_D}$}
\includegraphics[width=6.5125cm,height=5cm]{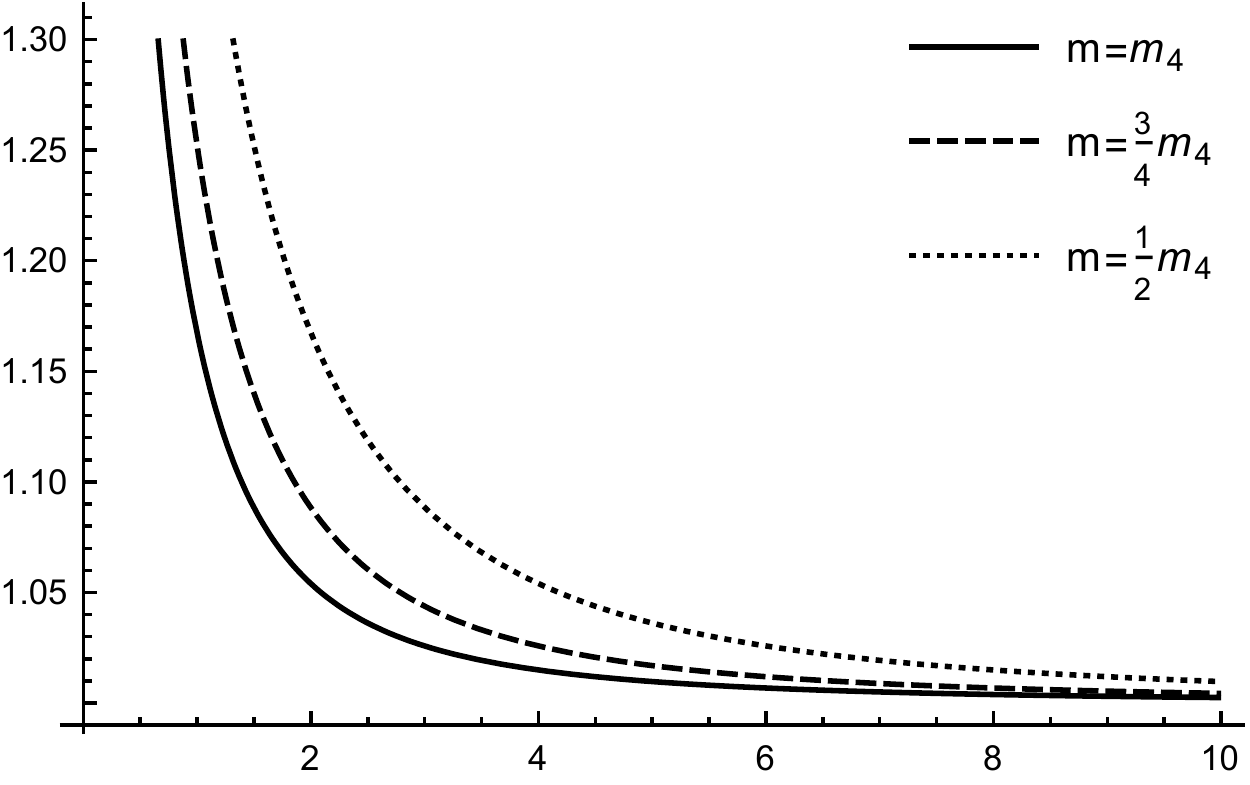}
$\quad$
\includegraphics[width=6.5125cm,height=5cm]{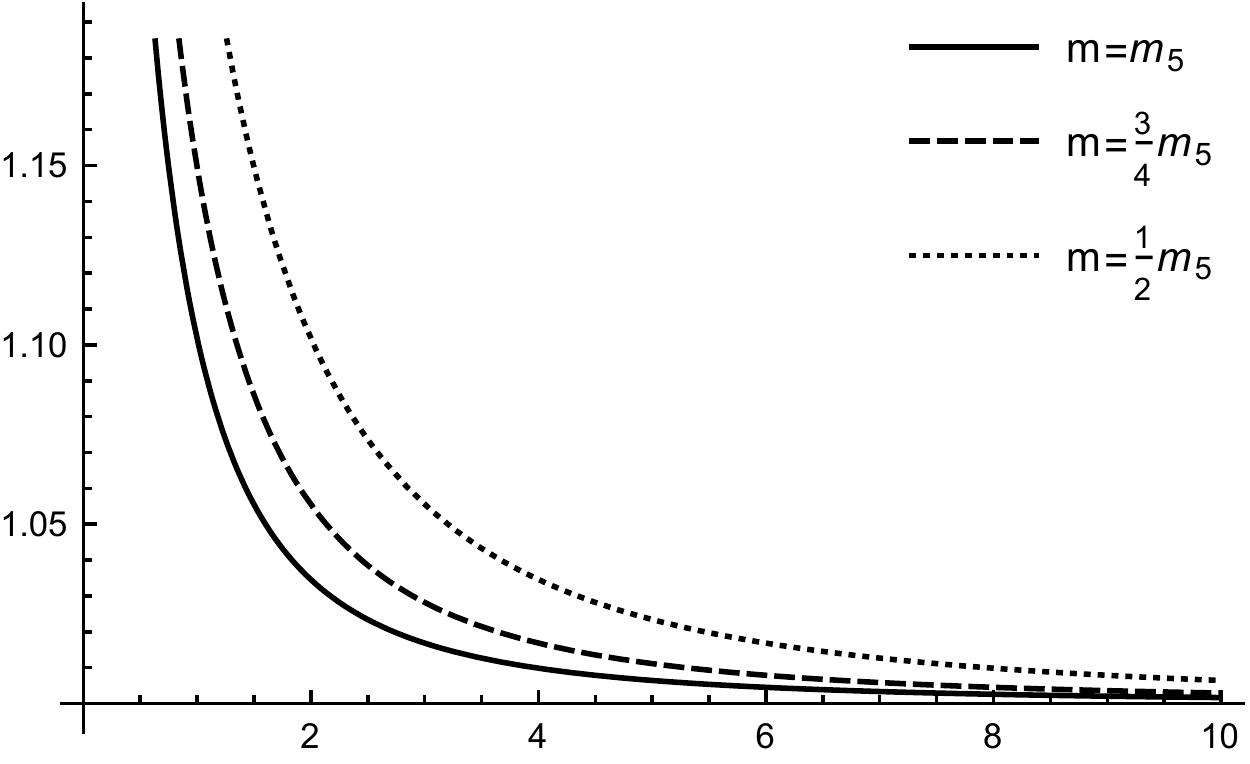}
{ $\ell/\ell_D$}
\\
\raisebox{4.5cm}{$\frac{\Delta{r}_{\rm H}}{R_D}$}
\includegraphics[width=6.5125cm,height=5cm]{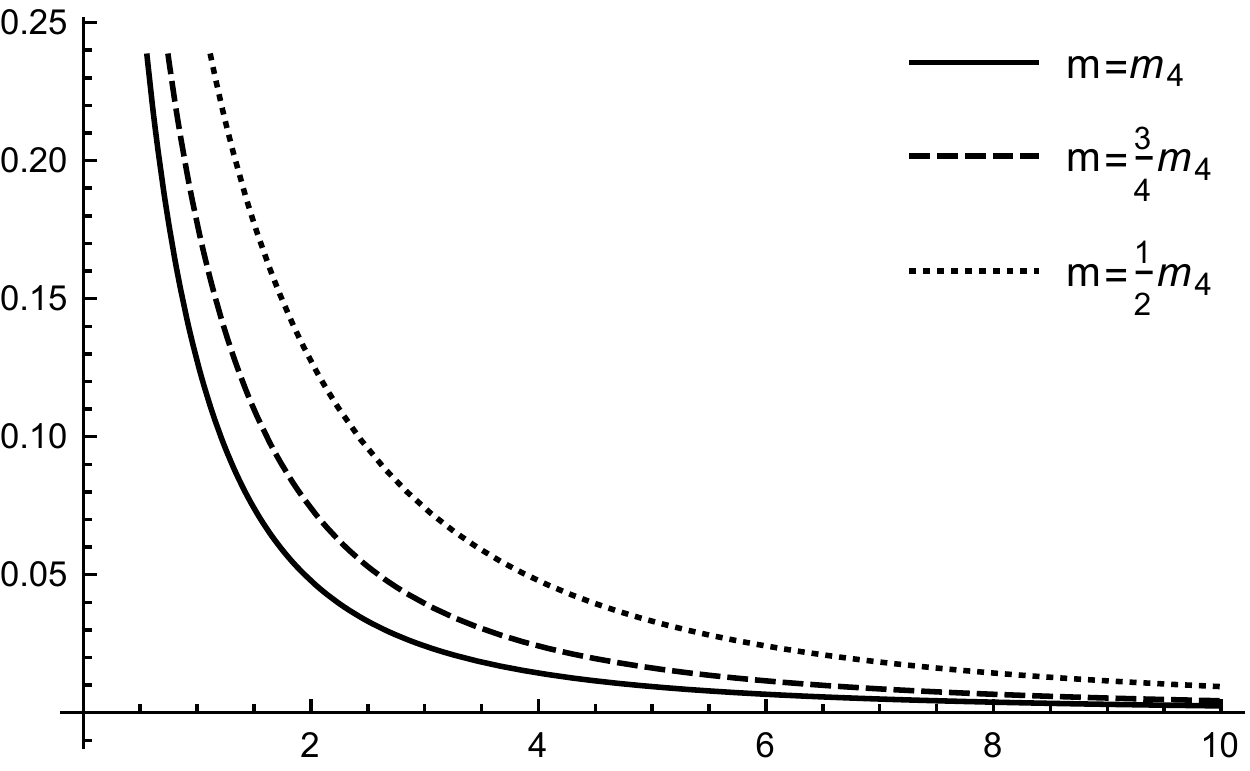}
$\quad$
\includegraphics[width=6.5125cm,height=5cm]{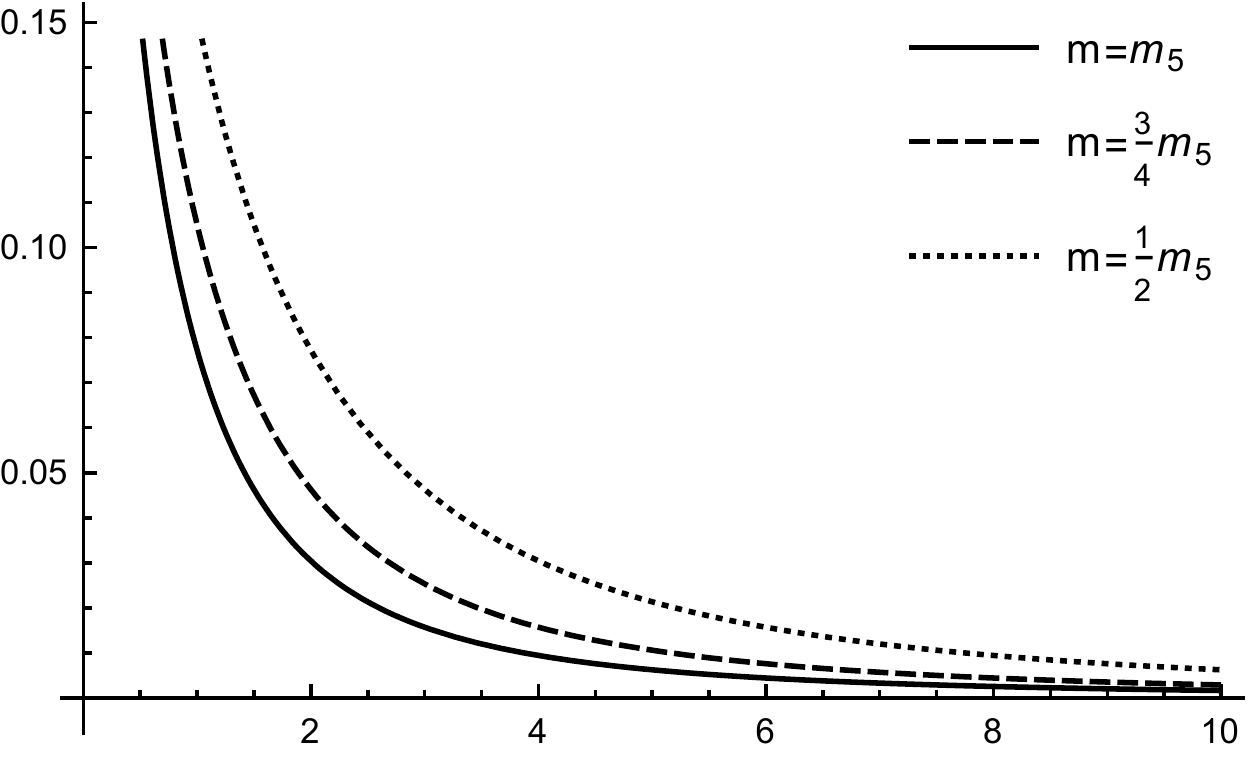}
{ $\ell/\ell_D$}
\\
$D=4$
\hspace{6cm}
$D=5$
\caption{Plots of $\expec{\hat{r}_{\rm H}}$ (upper panel) and $\Delta \rh$ (lower panels)
as functions of $\ell/\ell_D$, for $D=4$ (left) and $D=5$ (right), with $m=m_D$
(solid line), $m=\frac{3}{4}\,m_D$ (dashed line) and $m=\frac{1}{2}\,m_D$ (dotted line).
\label{expecD}}
\end{figure}

\begin{figure}[h!]
\centering
\raisebox{4.5cm}{$\frac{\expec{\hat{r}_{\rm H}}}{R_1}$}
\includegraphics[width=6.5125cm,height=5cm]{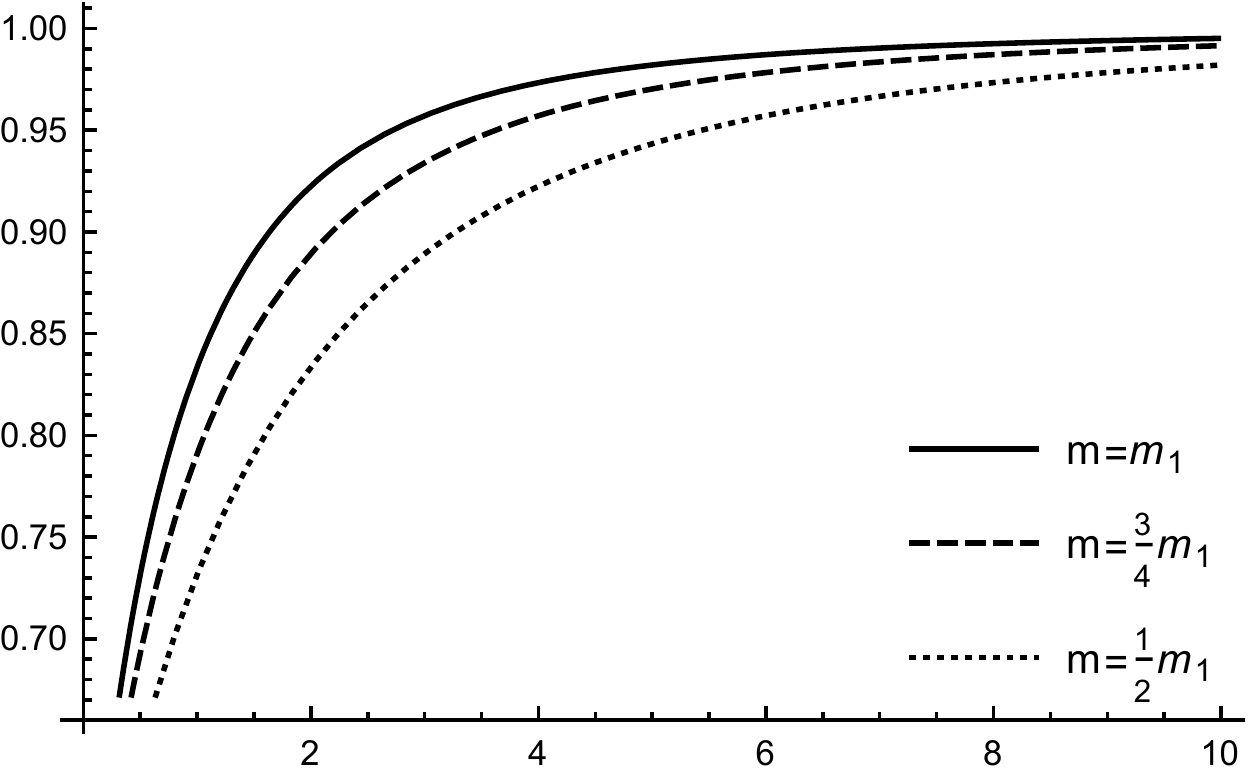}
$\quad$
\raisebox{4.5cm}{$\frac{\Delta{r}_{\rm H}}{R_1}$}
\includegraphics[width=6.5125cm,height=5cm]{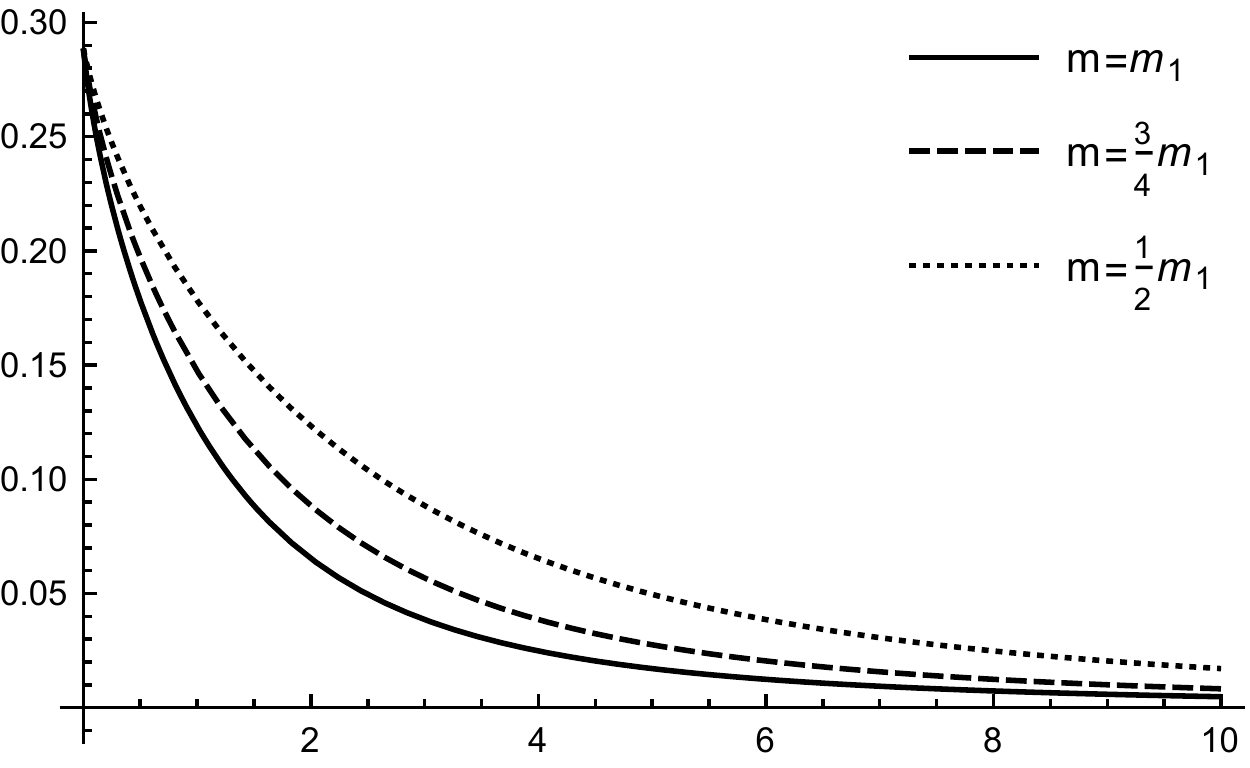}
{ $\ell/\ell_1$}
\caption{Plot of $\expec{\hat{r}_{\rm H}}$ and $\Delta \rh$
as functions of $\ell$ for $D=1$.
\label{expec1}}
\end{figure}

\setcounter{equation}{0}
We now derive the uncertainties in expectation values of quantities of relevance
in this framework, and as a result derive the form of the GUP.
Given the HWF~\eqref{HWF}, the expectation
value of an operator $\hat{O}_{\rm H}$ is obtained from
\be
\expec{\hat{O}_{\rm H}}
=
\Omega_{D-1} 
\int_0^\infty{\psi^*_{\rm H}(\rh) \, \hat{O}_{\rm H} \, \psi_{\rm H}(\rh)\,\rh^{D-1}\,\d \rh}
\ .
\ee
A straightforward example is given by
\be
\expec{\hat{r}_{\rm H}}
=
\frac{\Gamma\left(\frac{D+1}{2D-4},\frac{m^2}{\Delta^2}\right)}
{\Gamma\left(\frac{D}{2D-4},\frac{m^2}{\Delta^2}\right)}
\, \left(\frac{\Delta}{m}\right)^{\frac{1}{D-2}} \, R_D
\ .
\ee
By writing
$
\Gamma(s,x)
=
x^s \, \E_{1-s}(x)
$,
where $E_n$ is the generalised exponential integral
\be
\E_n(x)
\,=\,
\int_1^\infty\frac{e^{-xt}}{t^n} \, \d t
\ ,
\ee
the above result reads
\be
\expec{\hat{r}_{\rm H}}
=
\frac{\E_{\frac{D-5}{2D-4}}(\frac{m^2}{\Delta^2})}
{\E_{\frac{D-4}{2D-4}}(\frac{m^2}{\Delta^2})} \, R_D
\ .
\label{expecRhD}
\ee
We likewise obtain
\be
\expec{\hat{r}_{\rm H}^2}
=
\frac{\E_{\frac{D-6}{2D-4}}(\frac{m^2}{\Delta^2})}
{\E_{\frac{D-4}{2D-4}}(\frac{m^2}{\Delta^2})} \, R_D^2
\ ,
\ee
and estimate the relative uncertainty in the horizon as
\be
\Delta \rh
=
\sqrt{\expec{\hat{r}_{\rm H}^2}-\expec{\hat{r}_{\rm H}}^2}
=
\sqrt{
\frac{\E_{\frac{D-6}{2D-4}}(\frac{m^2}{\Delta^2})}
{\E_{\frac{D-4}{2D-4}}(\frac{m^2}{\Delta^2})} -
\left(\frac{\E_{\frac{D-5}{2D-4}}(\frac{m^2}{\Delta^2})}
{\E_{\frac{D-4}{2D-4}}(\frac{m^2}{\Delta^2})}\right)^2 } \, R_D
\label{deltarhD}
\ .
\ee
Fig.~\ref{expecD} shows the plots of~\eqref{expecRhD} and~\eqref{deltarhD}
for $D>3$ as functions of $\ell/\ell_D$, since
\be
\frac{m}{\Delta}\,=\,
\frac{\ell \, m}{\ell_D \, m_D}
\,\varpropto\,
\frac{\ell}{\ell_D}
\ .
\ee
It is also trivial to see that, for $\ell\gg\ell_D$, we recover the expected classical
results
\be
\expec{\hat{r}_{\rm H}}
\simeq
R_D
\ ,
\qquad
\Delta\rh
\simeq
0
\ .
\ee
The above expressions~\eqref{expecRhD} and~\eqref{deltarhD} also hold in $D=1$,
and are displayed in Fig.~\ref{expec1}.
\par
\begin{figure}[t!]
\centering
\raisebox{3.3cm}{$\frac{\Delta r}{\ell_D}$}
\includegraphics[width=4.5cm,height=3.625cm]{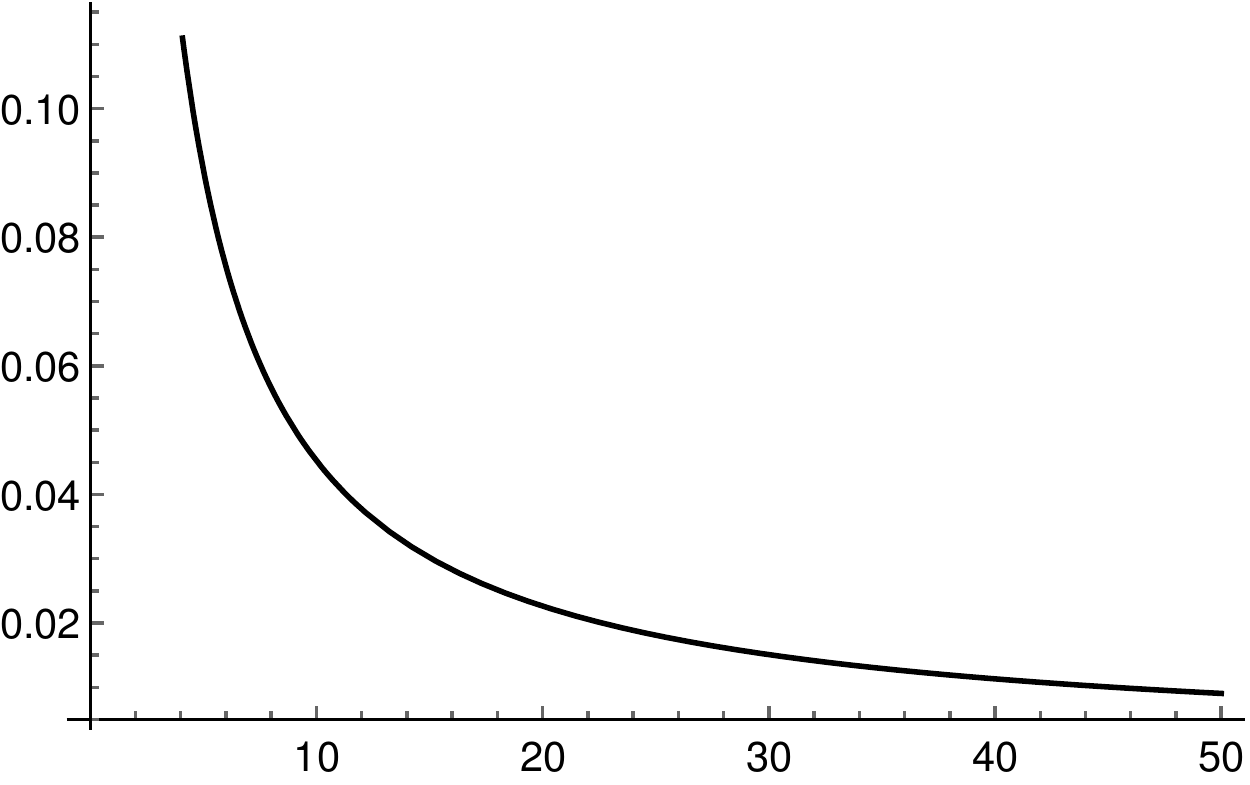}
$\quad$
\includegraphics[width=4.5cm,height=3.625cm]{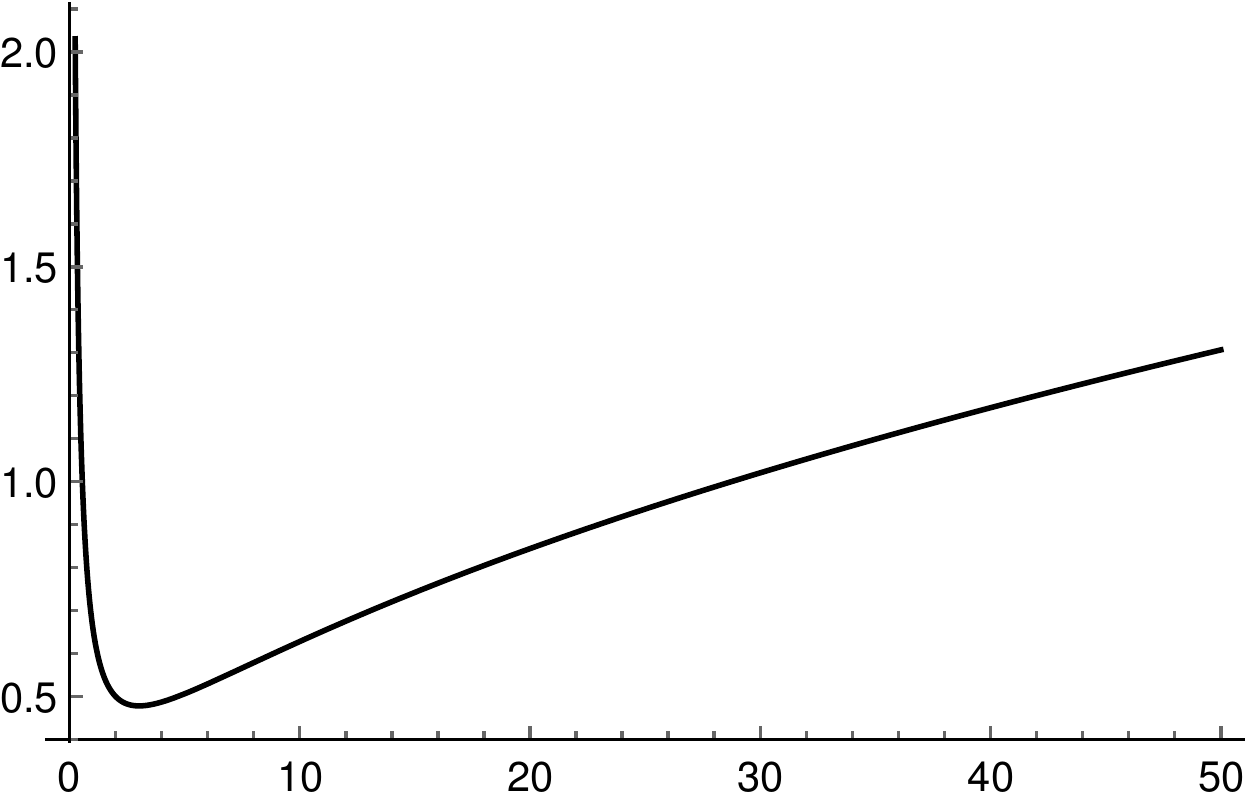}
$\quad$
\includegraphics[width=4.5cm,height=3.625cm]{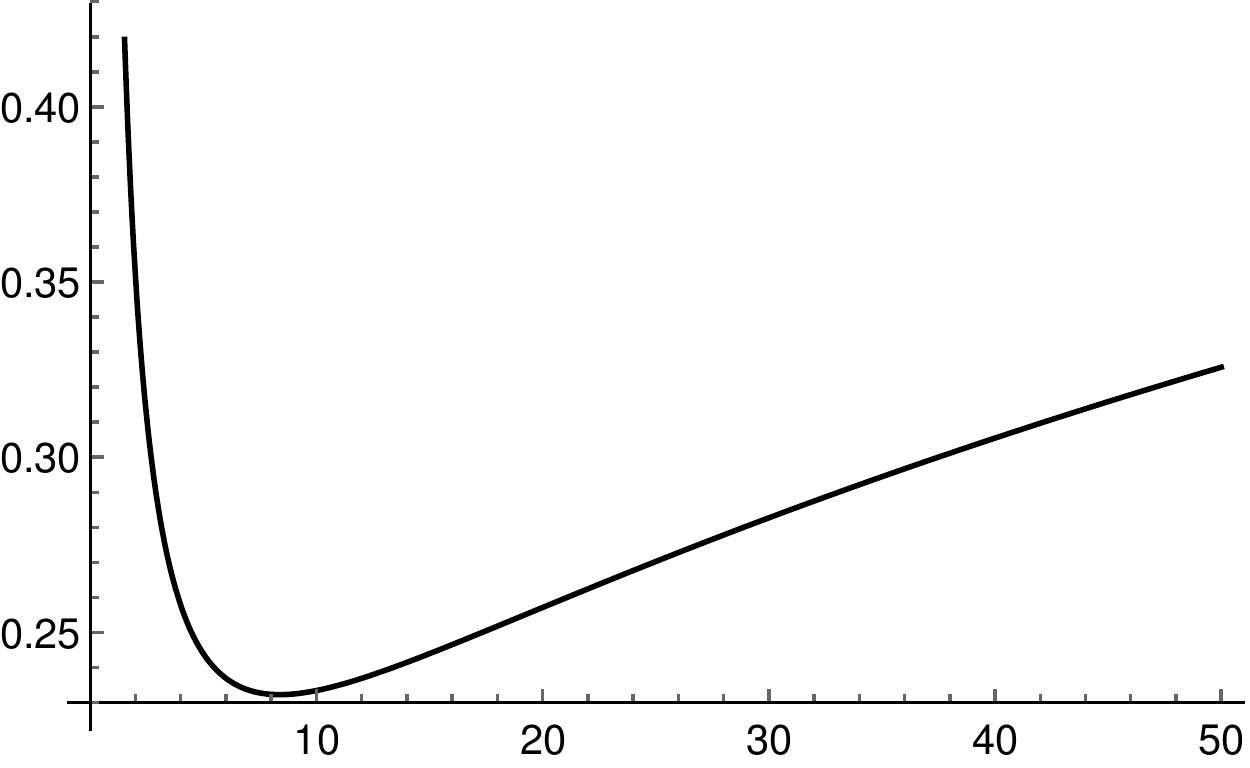}
$\frac{\Delta p}{m_D}$
\\
$D=1$
\hspace{4cm}
$D=4$
\hspace{4cm}
$D=5$
\caption{Plots of $\Delta r/\ell_D$ as function of $\Delta p/m_D$ for $D=1$, $4$ and $5$
 and $\alpha=1$.
\label{delta-r}}
\end{figure}
\begin{figure}[h!]
\centering
\raisebox{3.3cm}{\tiny ${\frac{M_D}{m_D}}$}
\includegraphics[width=4.7125cm,height=3.625cm]{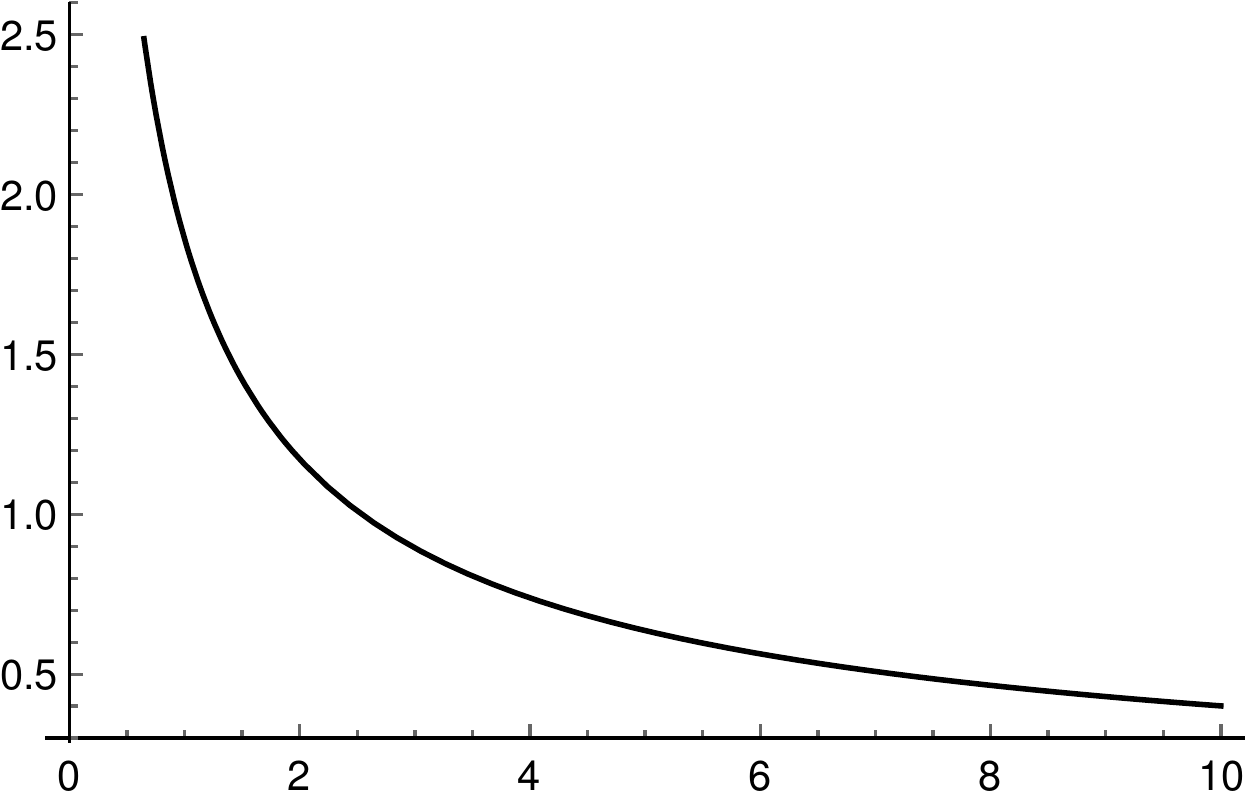}
$\quad$
\includegraphics[width=4.7125cm,height=3.625cm]{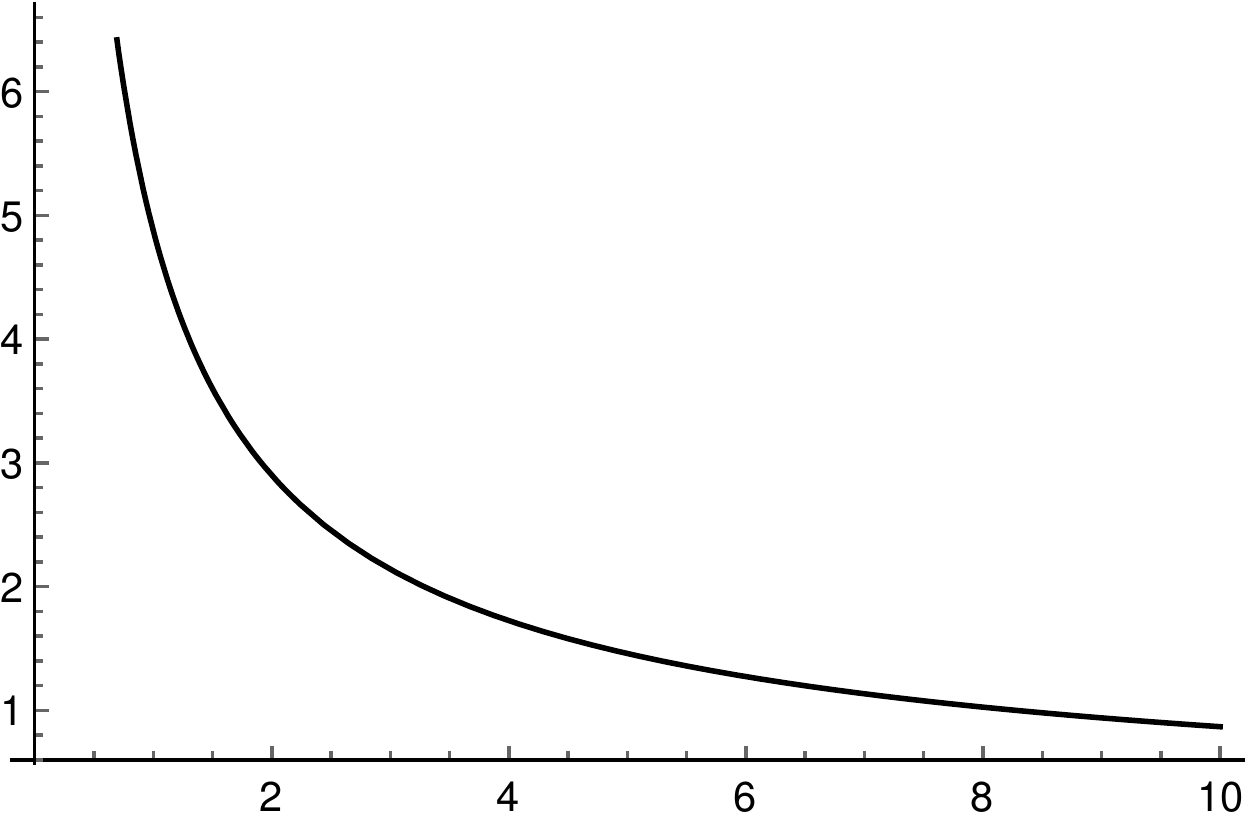}
$\alpha$
\\
$D=4$
\hspace{4cm}
$D=5$
\caption{Minimum scale $M_D$ as function of the parameter $\alpha$
for $D=4$ and $5$.
\label{scM}}
\end{figure}
\begin{figure}[h!]
\centering
\raisebox{3.3cm}{\tiny ${\frac{L_D}{\ell_D}}$}
\includegraphics[width=4.7125cm,height=3.625cm]{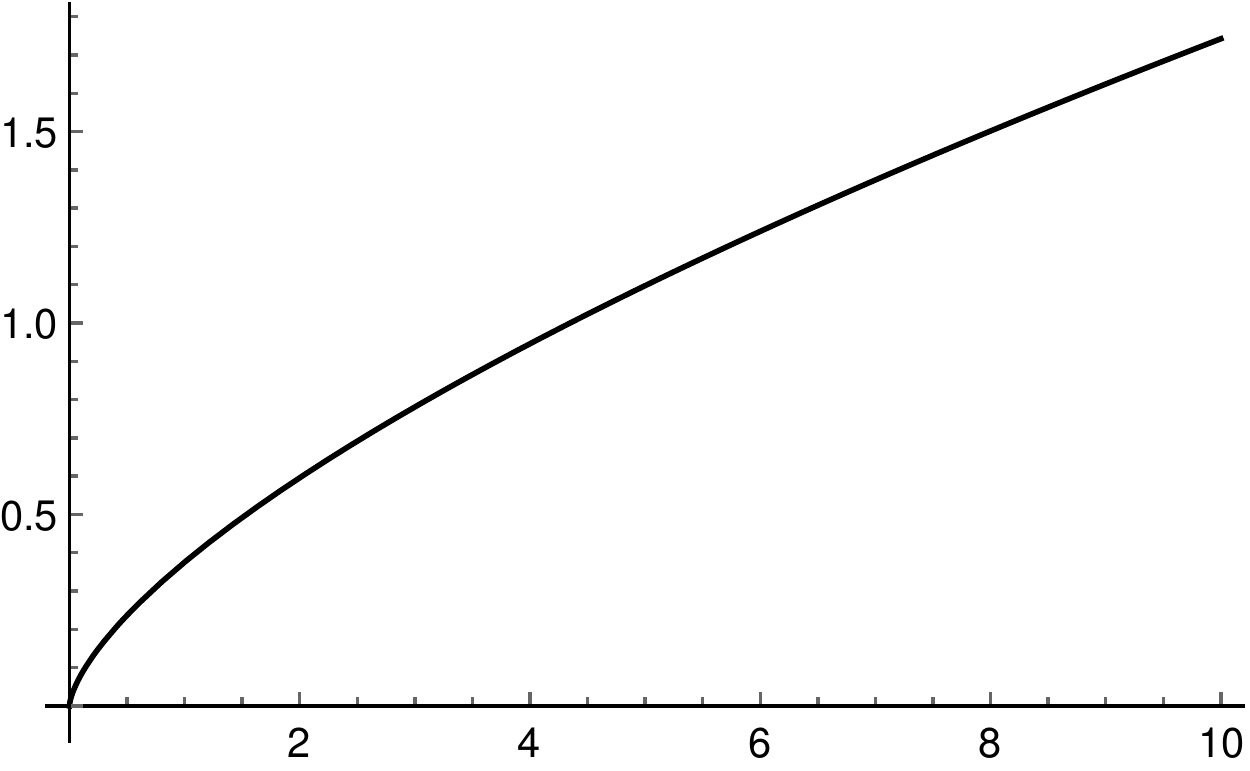}
$\quad$
\includegraphics[width=4.7125cm,height=3.625cm]{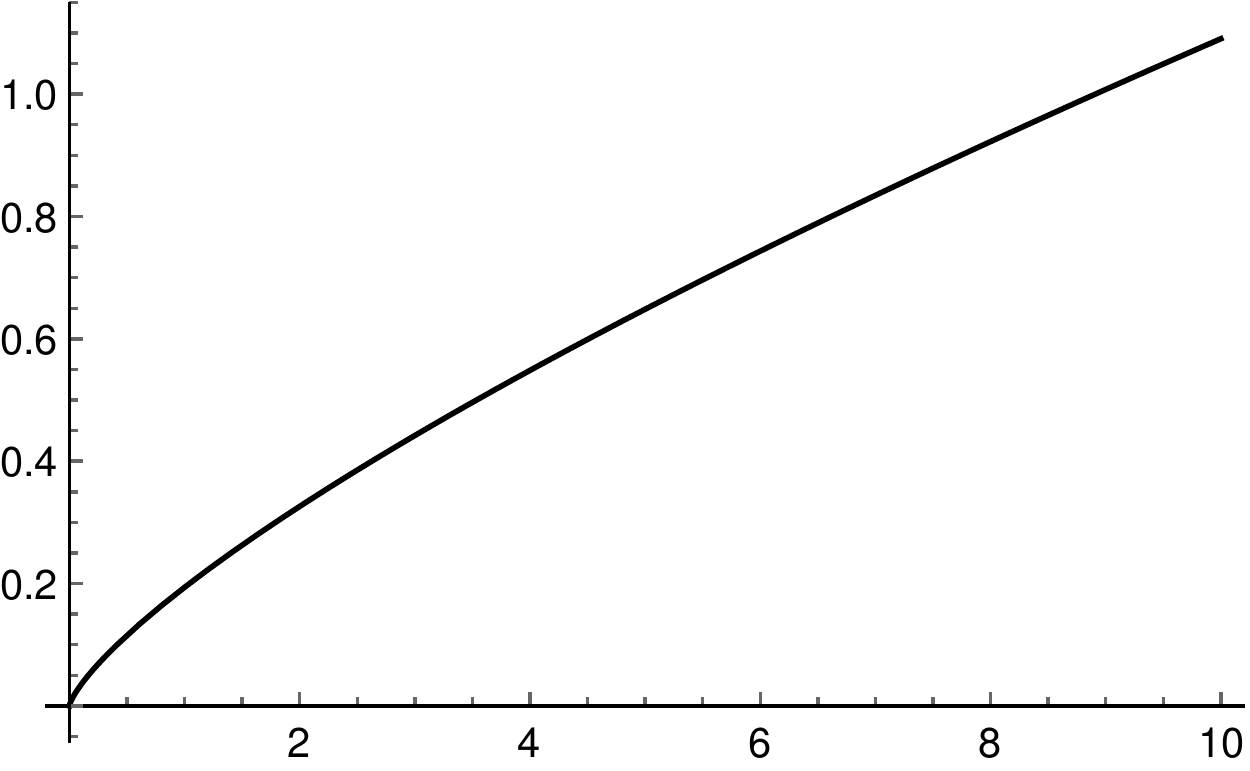}
$\alpha$
\\
$D=4$
\hspace{4cm}
$D=5$
\caption{Minimum scale $L_D$ as function of the parameter $\alpha$
for $D=4$ and $5$.
\label{scL}}
\end{figure}

The GUP follows by linearly combining the usual Heisenberg
uncertainty with the uncertainty in the horizon size, 
\be
\Delta r
=
\Delta r_{\rm QM} + \alpha \,\Delta \rh
\ ,
\ee
where $\alpha$ is a dimensionless coefficient that one could try to set
experimentally.
We can compute the Heisenberg part starting from the state~\eqref{Gauss},
that is 
\be
\expec{\hat{r}^n}
=
\Omega_{D-1} \int_0^\infty \tilde{\psi}^*(r) \, \hat{r}^n \, \tilde{\psi}(r) \, r^{D-1} \, \d r
=
\frac{\Gamma\left(\frac{D+n}{2}\right)}{\Gamma\left(\frac{D}{2}\right)} \, \ell^n
\ .
\ee
Using $\Gamma(z+1)=z\,\Gamma(z)$, this yields
\be
\expec{\hat{r}}
=
\frac{2^{1-D}\sqrt{\pi}\, (D-1)!}{\Gamma\left(\frac{D}{2}\right)^2} \, \ell
\ ,
\ee
and
\be
\expec{\hat{r}^2}
=
\frac{D}{2} \, \ell^2
\ ,
\ee
so that
\be
\Delta r_{\rm QM}
=
\sqrt{\frac{D}{2}
-\left(\frac{2^{1-D}\sqrt{\pi}}{\Gamma\left(\frac{D}{2}\right)^2} \, (D-1)!\right)^2} \, \ell
=
A_D \, \ell
\label{Drell}
\ .
\ee 
Using instead the state~\eqref{momGauss} in momentum space, the same procedure
yields
\be
\Delta p
=
A_D \, \Delta
=
A_D \, \frac{m_D \, \ell_D}{\ell}
\label{Deltap}
\ .
\ee
Expressing $\ell$ and $\Delta$ from the above equation as functions
of $\Delta p$, we have
\be
\frac{\Delta r}{\ell_D}
\!\!&=&\!\!
A_D^2\,\frac{m_D}{\Delta p}+
\alpha \,
\sqrt{
\frac{\E_{\frac{D-6}{2D-4}}\left(\frac{A_D^2 m^2}{(\Delta p)^2}\right)}
{\E_{\frac{D-4}{2D-4}}\left(\frac{A_D^2 m^2}{(\Delta p)^2}\right)} -
\left[\frac{\E_{\frac{D-5}{2D-4}}\left(\frac{A_D^2 m^2}{(\Delta p)^2}\right)}
{\E_{\frac{D-4}{2D-4}}\left(\frac{A_D^2 m^2}{(\Delta p)^2}\right)}\right]^2 } 
\, \left(\frac{2}{|D-2|}\, \frac{m}{m_D}\right)^\frac{1}{D-2} 
\ ,
\label{GUP}
\ee
which is rather cumbersome.
A straightforward simplification occurs by setting width of the wave-packet 
equal to the Compton length, $\ell=\lambda_m$, so that, from Eqs.~\eqref{lCompt}
and~\eqref{Deltap},
\be
m
=
\Delta
=
\frac{\Delta p}{A_D}
\ee
and the GUP then reads
\be
\frac{\Delta r}{\ell_D}
\!\!&=&\!\!
A_D^2\frac{m_D}{\Delta p}+
\alpha \,
\sqrt{
\frac{\E_{\frac{D-6}{2D-4}}(1)}{\E_{\frac{D-4}{2D-4}}(1)} -
\left(\frac{\E_{\frac{D-5}{2D-4}}(1)}{\E_{\frac{D-4}{2D-4}}(1)}\right)^2 } 
\left(\frac{2}{|D-2|}\, \frac{\Delta p}{A_D m_D}\right)^\frac{1}{D-2} 
\nonumber
\\
\!\!&=&\!\!
\frac{C_{\rm QM}}{\Delta p}
+C_{\rm H}\,\Delta p^{\frac{1}{D-2}}
\ ,
\label{GUPCompton}
\ee
where $C_{\rm QM}$ and $C_{\rm H}$ are constants (independent of $\Delta p$).
Fig.~\ref{delta-r} shows $\Delta r$ for different spatial dimensions,
setting $\alpha=1$ for simplicity.
In all the higher-dimensional cases, we obtain the same qualitative behaviour, 
with a minimum length uncertainty $L_D$
\be
L_D
=
\ell_D\left(\frac{D-1}{D-2}\right) \, (2A_D)^{\frac{1}{D-1}}
\left(\alpha\sqrt{\frac{\E_{\frac{D-6}{2D-4}}(1)}{\E_{\frac{D-4}{2D-4}}(1)} -
\left(\frac{\E_{\frac{D-5}{2D-4}}(1)}
{\E_{\frac{D-4}{2D-4}}(1)}\right)^2 }\right)^{\frac{D-2}{D-1}}
\label{MinL}
\ee
corresponding to an energy scale $M_D$, satisfying
\be
M_D
=
m_D \, \frac{(D-2)}{2^{\frac{1}{D-1}}}\,\left[
\alpha\sqrt{\frac{\E_{\frac{D-6}{2D-4}}(1)}{\E_{\frac{D-4}{2D-4}}(1)} -
\left(\frac{\E_{\frac{D-5}{2D-4}}(1)}
{\E_{\frac{D-4}{2D-4}}(1)}\right)^2 }\right]^{\frac{2-D}{D-1}} \, A_D^{\frac{2D-3}{D-1}}
\ .
\label{MinM}
\ee
The impact of $\alpha$ on this minimum length is then shown in Fig.~\ref{scM},
where we plot the scale $M_D$ corresponding to the minimum $L_D$
as a function of this parameter, and in Fig.~\ref{scL}, where
we plot directly $L_D$.
For all values of $D$ considered here, assuming $M_D\simeq m_D$ favours large
values of $\alpha$, whereas requiring $L_D\simeq \ell_D$ would favour small
values of $\alpha$.

\par
If we now apply the above result~\eqref{GUPCompton} in $D=1$, we see that
the complete expression is of the usual Heisenberg form,
\be
\frac{\Delta r}{\ell_1}
&\!\!=\!\!&
A_1^2 \, \frac{m_1}{\Delta p}
+\frac{\alpha}{2}\sqrt{
\frac{\E_{\frac{5}{2}}(1)}{\E_{\frac{3}{2}}(1)} -
\left(\frac{\E_{2}(1)}
{\E_{\frac{3}{2}}(1)}\right)^2 } \, \frac{A_1 m_1}{\Delta p}
\nonumber
\\
&\!\!=\!\!&
\frac{C_{\rm QM}+C_{\rm H}}{\Delta p}
\ ,
\ee
and there is no minimum length for any mass-scale.
This conclusion can also be inferred by the fact that Eqs.~\eqref{MinL}
and~\eqref{MinM} are not well defined in $D=1$.
\section{Conclusions}
\setcounter{equation}{0}
\label{Conclusions}
In this paper we extended the results of Refs.~\cite{Casadio:2013tma,Casadio:2014twa}
by embedding a massive source in a $(1+D)$-dimensional space-time.
Shaping the wave packet with a Gaussian distribution, we computed its related
horizon wave function and derived the probability $P_{\rm BH}$ that this massive source
be inside its own horizon, which characterises a (quantum) black hole.
\par
The higher-dimensional cases $D>3$ look qualitatively very similar
to the standard $(3+1)$-scenario, with a probability $P_{\rm BH}$ of similar shape,
and a related GUP leading to the existence of a minimum length scale.
However, one of the main results is that the probability $P_{\rm BH}$
for fixed mass decreases for increasing $D>3$.
In fact, for $m\simeq m_D$, one has $P_{\rm BH}\simeq 0.14$ for $D=5$,
which further decreases to $P_{\rm BH}\simeq 3\times 10^{-3}$ for $D=9$.
This implies that, although the fundamental scale $m_D$ could be smaller
for larger $D$, one must still reach an energy scale significantly larger than $m_D$
in order to produce a black hole.
It is clear that this should have a strong impact on the estimates of the number of 
black holes produced in colliders which are based on models with extra spatial dimensions,
and, conversely, on the bounds on extra-dimensional parameters obtained from the lack of
observation of these objects.
\par
We also note that the parameters of the source have not
been restricted to satisfy Buchdahl's inequality~\cite{buchdahl}, which is essentially
the condition for a source not to be a black hole in classical general relativity. 
Since the HWF is explicitly devised to include quantum effects, such inequality cannot
be assumed.
We can however see that one indeed finds a large probability that the object is a black
hole when Buchdahl's inequality in $D=3$ is violated (that is, for $m\gtrsim\mpl$).
For $D>3$, our conclusion can therefore be viewed as indicating a correction to 
Buchdahl's inequality, since the ratio $m/m_D$ must be larger for larger
$D$ in order to have a (significant probability to form a) black hole.

\par
The effective $(1+1)$-dimensional scenario differs from the higher-dimensional models.
In the latter case, the black hole probability increases as the energy of the system
(\emph{i.e.}~the mass of the particle) grows above the relevant ``gravitational energy scale''
$m_D$.
On the contrary, in $D=1$, black holes with masses far below $m_1$ are more likely,
but the maximum probability never exceeds $P_{\rm BH}\simeq 0.5$ regardless of the mass.
These features support the claim that two-dimensional black holes are purely quantum objects,
and are particularly important for the sub-Planckian regime of lower dimensional theories, 
where an effective dimensional reduction is expected~\cite{Mureika:2012fq}.
Moreover, the related GUP further supports the above arguments, since in $D=1$
there is no minimum length or mass.
\par
We would like to conclude by remarking the fact that the present analysis does not consider
the time evolution of the system, as the particle is taken at a fixed instant of time,
which is therefore left for future extensions.
\subsection*{Acknowledgments}
R.~C.~and A.~G.~are partly supported by INFN grant FLAG.
R.~T.~C.~is supported by CAPES and PDSE.~  J.~M.~was supported by
a Continuing Faculty Grant from the Frank R. Seaver College of
Science and Engineering at LMU.
\end{document}